\documentclass[twocolumn, twocolappendix]{aastex631}
\usepackage{gensymb}
\usepackage{amsmath}

\submitjournal{ApJ}

\graphicspath{{./}{figures/}}

\begin{document}

\title{A New Constraint on the Relative Disorder of Magnetic Fields between Neutral \\ Interstellar Medium Phases}


\author[0000-0002-2679-4609]{Minjie Lei}
\affiliation{Department of Physics, Stanford University, Stanford, CA 94305, USA}
\affiliation{Kavli Institute for Particle Astrophysics \& Cosmology, P.O. Box 2450, Stanford University, Stanford, CA 94305, USA}

\author[0000-0002-7633-3376]{S. E. Clark}
\affiliation{Department of Physics, Stanford University, Stanford, CA 94305, USA}
\affiliation{Kavli Institute for Particle Astrophysics \& Cosmology, P.O. Box 2450, Stanford University, Stanford, CA 94305, USA}

\begin{abstract}

Utilizing Planck polarized dust emission maps at 353 GHz and large-area maps of the neutral hydrogen (\ion{H}{1}) cold neutral medium (CNM) fraction ($f_\mathrm{CNM}$), we investigate the relationship between dust polarization fraction ($p_{353}$) and $f_\mathrm{CNM}$ in the diffuse high latitude ($\left|b\right|>30\degree$) sky. We find that the correlation between $p_{353}$ and $f_\mathrm{CNM}$ is qualitatively distinct from the $p_{353}$-\ion{H}{1} column density ($N_\mathrm{H\,I}$) relationship. At low column densities ($N_\mathrm{H\,I}<4\times10^{20}~\mathrm{cm^{-2}}$) where $p_{353}$ and $N_\mathrm{H\,I}$ are uncorrelated, there is a strong positive $p_{353}$-$f_\mathrm{CNM}$ correlation. We fit the $p_{353}$-$f_{\rm CNM}$ correlation with data-driven models to constrain the degree of magnetic field disorder between phases along the line-of-sight. We argue that an increased magnetic field disorder in the warm neutral medium (WNM) relative to the CNM best explains the positive $p_{353}$-$f_\mathrm{CNM}$ correlation in diffuse regions. Modeling the CNM-associated dust column as being maximally polarized, with a polarization fraction $p_{\rm CNM} \sim$ 0.2, we find that the best-fit mean polarization fraction in the WNM-associated dust column is 0.22$p_{\rm CNM}$. The model further suggests that a significant $f_{\rm CNM}$-correlated fraction of the non-CNM column (an additional ~18.4\% of the \ion{H}{1} mass on average) is also more magnetically ordered, and we speculate that the additional column is associated with the unstable medium (UNM). Our results constitute a {new} large-area constraint on the average relative disorder of magnetic fields between the neutral phases of the ISM, and are consistent with the physical picture of a more magnetically aligned CNM column forming out of a disordered WNM. 

\end{abstract}

\keywords{Interstellar medium (847) --- Cold neutral medium(266) --- Interstellar magnetic fields(845) --- Interstellar dust(836)}


\section{Introduction} \label{sec:intro}
Magnetic fields permeate the diffuse interstellar medium (ISM) and play an important role in astrophysical processes across our Galaxy, including gas dynamics, cosmic ray transport, molecular cloud and star formation \citep{lselvon11-cr, crutcher12-mc}. However, our current picture of how the structure of the interstellar magnetic field varies across the complex, multiphase ISM environment is poorly constrained. Observational tracers tend to only probe partial projections of the full 3D magnetic field in specific phases, and are often limited to sightline-averaged properties \citep[e.g.,][]{ferriere01-rv, han17-rv}. A complete understanding of magnetic fields in the ISM is an important and challenging problem that requires piecing together different tracers of magnetic fields and ISM phases. 

The ISM can be generally organized into ionized, neutral, and molecular phases across a wide range of temperatures and densities. The neutral ISM as traced by \ion{H}{1} is further composed of the cold neutral medium (CNM), the warm neutral medium (WNM), and the thermally unstable medium (UNM) \citep{field69-ph, wolfire03-na, kalberla09-hi}. One important tracer of magnetic fields in the neutral medium is polarized thermal dust emission. Spinning dust grains preferentially align with their major axes perpendicular to the local magnetic field \citep{andersson15-re}. The measurement of polarized dust emission traces the component of the magnetic field projected onto the plane of the sky (POS) in a weighted integral over {dust density} along the line of sight (LOS). The statistical properties of polarized dust emission maps have been used extensively to study the structure of the interstellar magnetic field \citep[e.g.,][]{planck14-pd, planck14-mh, planck18-cr, Fissel16-bd, chen19-bd, sullivan21-bd, hoang22-bl}. 

Recent work has also demonstrated that \ion{H}{1} gas in the diffuse, neutral ISM is organized into linear filamentary structures that align with the orientation of magnetic fields as traced by dust polarization \citep{clark15-nh, kalberla16-ah, clark18-np}. Furthermore, these \ion{H}{1} filaments have been shown to be preferentially associated with the cold neutral medium (CNM) \citep{mcclureGriffiths06-ab, clark14-ma, kalberla18-cw, clark19-pn, peek19-sc, murray20-cn}. \citet{clark19-hs} constructed \ion{H}{1}-based Stokes I, Q, and U maps with a Stokes polarization angle determined from the orientation of filaments in \ion{H}{1} channel maps, and showed that they are well-correlated with polarized dust emission. However, the polarized dust emission traces a weighted integral of the total dust column along the LOS, while the \ion{H}{1}-based polarization is determined by the orientation of primarily CNM structures. The strong correlation of polarized dust emission with a quantity constructed from the orientation of primarily CNM structures implies that the mean magnetic field orientation in the CNM is similar to the mean magnetic field orientation in the WNM. A study of the diffuse field targeted by the BICEP/Keck experiment found that measured dust polarization properties were consistent between the total polarized dust emission and the \ion{H}{1}-filament-correlated component of the polarized dust \citep{bk22-cd}. We are thus motivated to ask whether the relationship between dust and \ion{H}{1} data can constrain the relative dispersion of the magnetic field angle in the dust associated with the WNM as compared to the CNM.  

Other analyses have compared dust polarization to gas column density. Using Planck observations at 353 GHz, \citet{planck14-pd} examined the variation of the dust polarization fraction $p_{353}$ with total gas column density $N_{\mathrm{H}}$ for the full sky excluding the inner Galactic plane. They observed a column-density dependent behavior for the correlation between $p_{353}$ and $N_{\mathrm{H}}$. At low column densities ($2\times10^{20}\ \mathrm{cm^{-2}} < N_{\mathrm{H}} < 10^{21}\ \mathrm{cm^{-2}}$), there is significant variation in the values of $p_{353}$ from consistent with zero to the maximum observed value for the full sky ($p_{max}\sim0.2$), and there is no observed correlation between $p_{353}$ and $N_{\mathrm{H}}$. At higher column densities ($10^{21}\ \mathrm{cm^{-2}} < N_{\mathrm{H}} < 2\times10^{22}\ \mathrm{cm^{-2}}$), $p_{353}$ shows a clear anticorrelation with $N_{\mathrm{H}}$, with $\left<p_{353}\right>$ decreasing to below 0.04 at $N_{\mathrm{H}}>10^{21}\ \mathrm{cm^{-2}}$. \citet{planck14-mh} found that the observed anticorrelation at high column densities can be reproduced in simulations of MHD turbulence assuming uniform dust properties. This implies that the variation in polarization fraction on scales probed by Planck data can be explained by the tangling of magnetic field structures along the LOS and within the beam up to $N_{\mathrm{H}} \sim 10^{22}$. 

The primary probe of \ion{H}{1} is the 21cm line from the hyperfine transition of ground-state neutral hydrogen. The \ion{H}{1} spin temperature can be directly constrained from absorption and emission data \citep{heiles03-ml, murray18-sp}. However absorption measurements are limited by the availability of background continuum sources, and it is challenging to separate ISM phases from emission-line data alone. Significant progress on deriving \ion{H}{1} phase properties from \ion{H}{1} emission has been made using statistical approaches such as Gaussian phase decomposition \citep{haud07-gd, kalberla18-cw, marchal19-rs, riener20-ag}, wavelet scattering transform \citep{lei22-st}, and convolution neural networks (CNNs) \citep{murray20-cn}. In this study, we make use of the CNN-predicted CNM mass fraction ($f_\mathrm{CNM}$) map from \citet{murray20-cn}, to extend the study of $p$-$N_{\mathrm{H}}$ statistics to phase-decomposed \ion{H}{1} properties. 

The extent to which the structure of the magnetic field varies across the boundaries of the ISM phases is largely unknown. Using state-of-the-art large-area phase decomposed maps, we study the relationship between magnetic field tracers and gas phase, and derive new limits on the relative disorder of magnetic fields between ISM phases in the diffuse medium. Previous work on constraining the variation of magnetic field structure across phases has mostly been confined to limited regions of the sky where a direct morphological connection between tracers can be found. For example, several studies examining a 64 $\mathrm{deg}^2$ field centered on the quasar 3C196 have reported a morphological correlation between LOFAR Faraday rotation data probing the magneto-ionic medium and \ion{H}{1} and dust data probing the neutral medium \citep{zaroubi15-lh, jelic18-lh, bracco20-lh}. Comparing radio polarization and \ion{H}{1} emission data, \citet{campbell22-cm} find evidence for an aligned magnetic field between the warm ionized medium and the CNM in a small patch of high-Galactic latitude sky, but a lack of widespread association between these tracers in the full high latitude Arecibo sky. {Modeling dust emission associated with different \ion{H}{1} phases as discrete layers, and fitting to Planck observations in the polar cap regions, \citet{ghosh17-ms} and \citet{adak20-dp} concluded that the magnetic field in the CNM is more turbulent than the UNM/WNM layers}. Here we propose and apply a new method to constrain the relative disorder of magnetic field structures between the cold and warm phases of the neutral ISM.

The rest of the paper is organized as follows. In Section \ref{sec:data}, we detail the dust and \ion{H}{1} data products used in this study. In Section \ref{sec:corr_res}, we examine the correlation between the dust polarization fraction $p_{353}$ and the CNM mass fraction $f_\mathrm{CNM}$ in the diffuse ISM. In Section \ref{sec:corr_interp}, we examine different possible contributions to a $p_{353}$-$f_\mathrm{CNM}$ relation, and rule out explanations other than a phase-dependent degree of magnetic field tangling. Our model of magnetic field relative disorder between the WNM and CNM is presented in Section \ref{sec:corr_model}, followed by a discussion and conclusions in Sections \ref{sec:discussion} and \ref{sec:conclusions}. Further technical discussions can be found in Appendix \ref{appx:galfa_v_hi4pi}.

\section{Data} \label{sec:data}

\subsection{Polarized Dust Emission} \label{subsec:pfrac}
For polarized dust emission, we use the 80\arcmin\ R3.00 353 GHz Stokes $I$, $Q$, and $U$ maps released by the \textit{Planck} collaboration. A main source of uncertainty in estimating polarization fraction $p_{353}$ is the zero-level correction of the Stokes $I$ map. Following \citet{planck18-cr}, we adopt a CIB monopole subtraction of 452 $\mu\mathrm{K_{CMB}}$ at 353 GHz, and add a fiducial Galactic offset correction of 63 $\mu\mathrm{K_{CMB}}$ \citep{planck18-cr}. \citet{planck18-cr} estimate the offset uncertainty to be 40 $\mu\mathrm{K_{CMB}}$, which affects the shape of the estimated $p_{353}$ distribution but not the correlation statistics investigated in this study. Finally, from the 80\arcmin, offset-corrected 353 GHz Stokes $I$, $Q$, and $U$ maps and their covariances, we compute the noise-debiased polarization fraction $p_{353}$ using the modified asymptotic estimator method introduced by \citet{plaszczynski14-db}. We further apply a signal-to-noise ratio (SNR) cut of $p_{353}/\sigma_{p_{353}}>3$ (retaining $\sim93\%$ of the full sky) to arrive at the final map used for our analysis. For all the subsequent data products employed in this work, we smooth them to 80\arcmin\ to match the resolution of the polarization fraction map, and convert them to HEALPix format pixelated with $N_\mathrm{side}=64$ \citep{healpy05-hp}.

\subsection{Neutral Hydrogen Emission} \label{subsec:hi_column}
We use \ion{H}{1} emission data from Data Release 2 (DR2) of the Galactic Arecibo L-band Feed Array Survey \citep[GALFA-\ion{H}{1};][]{peek18-ga}. GALFA-\ion{H}{1} is a high angular (4\arcmin) and spectral (0.184 km/s) resolution survey that covers $\sim 32\%$ of the sky from Dec. $-1\degree17\arcmin$ to Dec. $+37\degree57\arcmin$ across all R.A. For analysis of total \ion{H}{1} column density, we utilize the publicly-available, stray-radiation-corrected column density maps from DR2 integrated over $|v_{\mathrm{LSR}}|\leq90$ km/s \citep{peek18-ga}. 

In addition to GALFA-\ion{H}{1}, we also employ data from the \ion{H}{1}4PI survey \citep{hi4pi16-rl}. \ion{H}{1}4PI combines the Effelsberg-Bonn \ion{H}{1} Survey of the northern sky \citep[EBHIS;][]{winkel16-eb} with the Galactic All Sky Survey using the Parkes radio telescope in the south \citep[GASS;][]{macclure-griffiths09-ga} to produce a full-sky \ion{H}{1} emission survey at 16.2\arcmin\ angular and 1.49 km/s spectral resolution. We similarly employ the total \ion{H}{1} column density maps from \ion{H}{1}4PI with stray-radiation correction applied \citep{hi4pi16-rl}. 

\subsection{Cold Neutral Medium Fraction} \label{subsec:fcnm}
To trace cold neutral medium content in the same footprints as the dust and \ion{H}{1} emission data, we utilize $f_{\mathrm{CNM}}$ maps produced using the convolutional neural net (CNN) model from \citet{murray20-cn}. The model is trained and tested on augmented, synthetic \ion{H}{1} emission spectra from 3D MHD simulations of the Galactic ISM \citep{kim13-si, kim14-si}. It is then validated against $f_{\mathrm{CNM}}$ measurements derived using available \ion{H}{1} absorption observations from 21-SPONGE \citep{murray18-sp} and the Millennium survey \citep{heiles03-ml}. The $\sigma_{f_{\mathrm{CNM}}}$ uncertainties on predicted $f_{\mathrm{CNM}}$ values are estimated by running multiple iterations of the model with different random initialization. We make use of the $f_{\mathrm{CNM}}$ maps produced by applying the model to GALFA-\ion{H}{1} and \ion{H}{1}4PI 21cm emission data respectively at high Galactic latitudes ($|b|>30\degree$). The GALFA-\ion{H}{1} $f_{\mathrm{CNM}}$ map is presented in \citet{murray20-cn}, while the \ion{H}{1}4PI version of the map is first described in \citet{hensley22-pa}. The two maps are shown to have excellent agreement in overlapping regions \citep{hensley22-pa}.  

\begin{figure*}[t]
\centering
\includegraphics[width=0.95\textwidth]{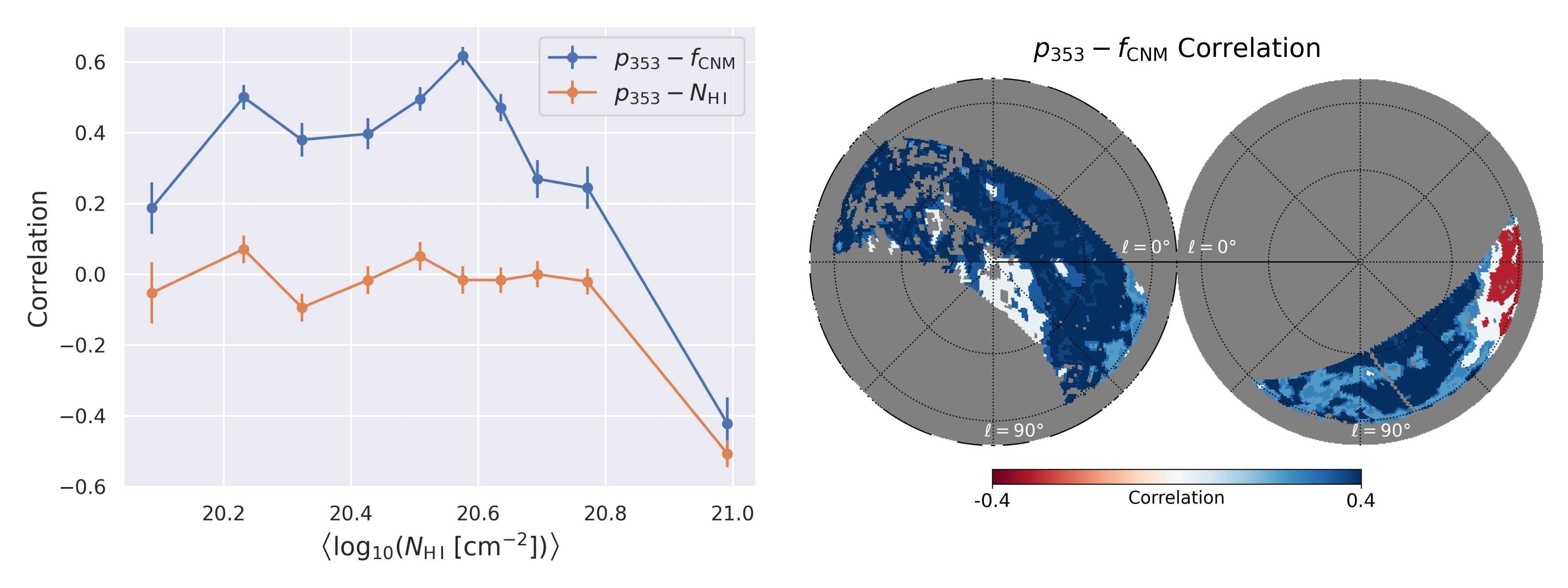}
\caption{The $p_{353}$-$f_{\mathrm{CNM}}$ and $p_{353}$-$N_{\mathrm{H\,I}}$ correlations in the high Galactic latitude ($|b|>30\degree$) GALFA-\ion{H}{1} sky. Left: Spearman correlation coefficients computed in bins of $N_{\mathrm{H\,I}}$, such that each bin covers an approximately equal sky area. The data used to compute these correlations are shown for selected bins in Figure \ref{fig:compare_col_den}. While both $f_{\mathrm{CNM}}$ and $N_{\mathrm{H\,I}}$ anticorrelate with $p_{353}$ at high column density ($\sim 10^{21}~\mathrm{cm}^{-2}$), a strong, consistent correlation between $f_{\mathrm{CNM}}$ and polarization fraction is observed at lower column density regimes where $p_{353}$-$N_{\mathrm{H\,I}}$ is compatible with no correlation. Right: Map of correlation between $p_{353}$ and $f_{\mathrm{CNM}}$, created by dividing the sky into 20 $N_{\mathrm{H\,I}}$ bins with approximately equal area, colored by the $p_{353}$-$f_{\mathrm{CNM}}$ correlation coefficient value in each bin. {The maps are shown in an orthographic projection centered on $(l, b)=(0, 90)$ (left) and $(l, b)=(0, -90)$ (right)}. The positively and negatively correlated regions tend to be fairly spatially coherent. }
\label{fig:p353_fcnm_corr}
\end{figure*}

\subsection{\ion{H}{1} Velocity Components along the LOS} \label{subsec:ncloud}
One potential contribution to the variation of the dust polarization fraction is depolarization due to the complexity of the dust distribution along the LOS. \citet{panopoulou20-nc} developed a method for quantifying the LOS complexity of \ion{H}{1} emission, which can be used as a proxy for the LOS distribution of the associated dust. By applying Gaussian decomposition to identify distinct emission components in the \ion{H}{1} spectra from the 
\ion{H}{1}4PI survey, the authors created maps of the number of clouds along the LOS. A wide Gaussian kernel bandwidth of 5~km/s is selected so that the components are effectively multi-phase clouds and not sensitive to narrower CNM emission features. The maps cover the area of the high Galactic latitude sky where the \ion{H}{1} column density is linearly correlated with far infrared dust emission, i.e. where $N_\mathrm{H\,I}<4\times10^{20}~\mathrm{cm^{-2}}$ \citep{lenz17-ln}. \citet{panopoulou20-nc} additionally defined a LOS complexity measure weighted by the column density of the components
\begin{equation} \label{eq:nc}
    \mathcal{N}_c = \sum_{i=1}^{N_\mathrm{clouds}}\frac{N_\mathrm{H\,I}^i}{N_\mathrm{H\,I}^\mathrm{max},}
\end{equation}
where $N_\mathrm{H\,I}^i$ is the $i$th cloud along the sightline, with $N_\mathrm{H\,I}^\mathrm{max}$ the maximum column density of the clouds identified for that sightline. In the case where one single cloud dominates the total column density of the sightline, $\mathcal{N}_c$ will be $\sim 1$. Meanwhile, a sightline with $n$ equal-column-density clouds will have $\mathcal{N}_c=n$. Higher $N_c$ is primarily driven by the presence of intermediate velocity clouds \citep{panopoulou20-nc}. 

In this paper, we employ $\mathcal{N}_c$ maps to quantify how much the variation of dust polarization fraction with \ion{H}{1} phase content is attributable to the complexity of the overall \ion{H}{1} distribution along the LOS.

\subsection{\ion{H}{1} Polarization Template} \label{subsec:hi_pol}
The morphology of \ion{H}{1} emission intensity structures encodes measurable information about \ion{H}{1} phases and the properties of the ambient magnetic field.
\citet{clark18-np} introduced a formalism to characterize the LOS tangling of the magnetic field from the orientation of \ion{H}{1} structures at different LOS velocities. From the linear intensity of \ion{H}{1} structures mapped using the Rolling Hough Transform \citep[RHT;][]{clark14-ma}, \citet{clark19-hs} constructed 3D (position-position-velocity) Stokes $I_{\rm H\,I}$, $Q_{\rm H\,I}$, and $U_{\rm H\,I}$ maps based on \ion{H}{1}4PI survey data. After integrating over the velocity dimension, the \ion{H}{1}-based polarization maps are well-correlated with the 353 GHz polarized dust emission maps \citep[see also][]{bk22-cd, halal23-fd}. 

In this work, we utilize the velocity-integrated, publicly available \citet{clark19-hs} \ion{H}{1}-based polarization maps, and derive a polarization fraction $p_\mathrm{H\,I}$ and polarization angle $\theta_{\rm H\,I}$ from the \ion{H}{1}-based Stokes parameters smoothed to the same resolution as the Planck 353 GHz map. We compare the $p_\mathrm{H\,I}$-$f_\mathrm{CNM}$ correlation to the $p_{353}$-$f_\mathrm{CNM}$ correlation and explore the possible implications for the variation of magnetic field structures across \ion{H}{1} phases.

\section{Dust Polarization Fraction Is Correlated with CNM Content} \label{sec:corr_res}

\begin{figure*}[t]
\centering
\includegraphics[width=0.95\textwidth]{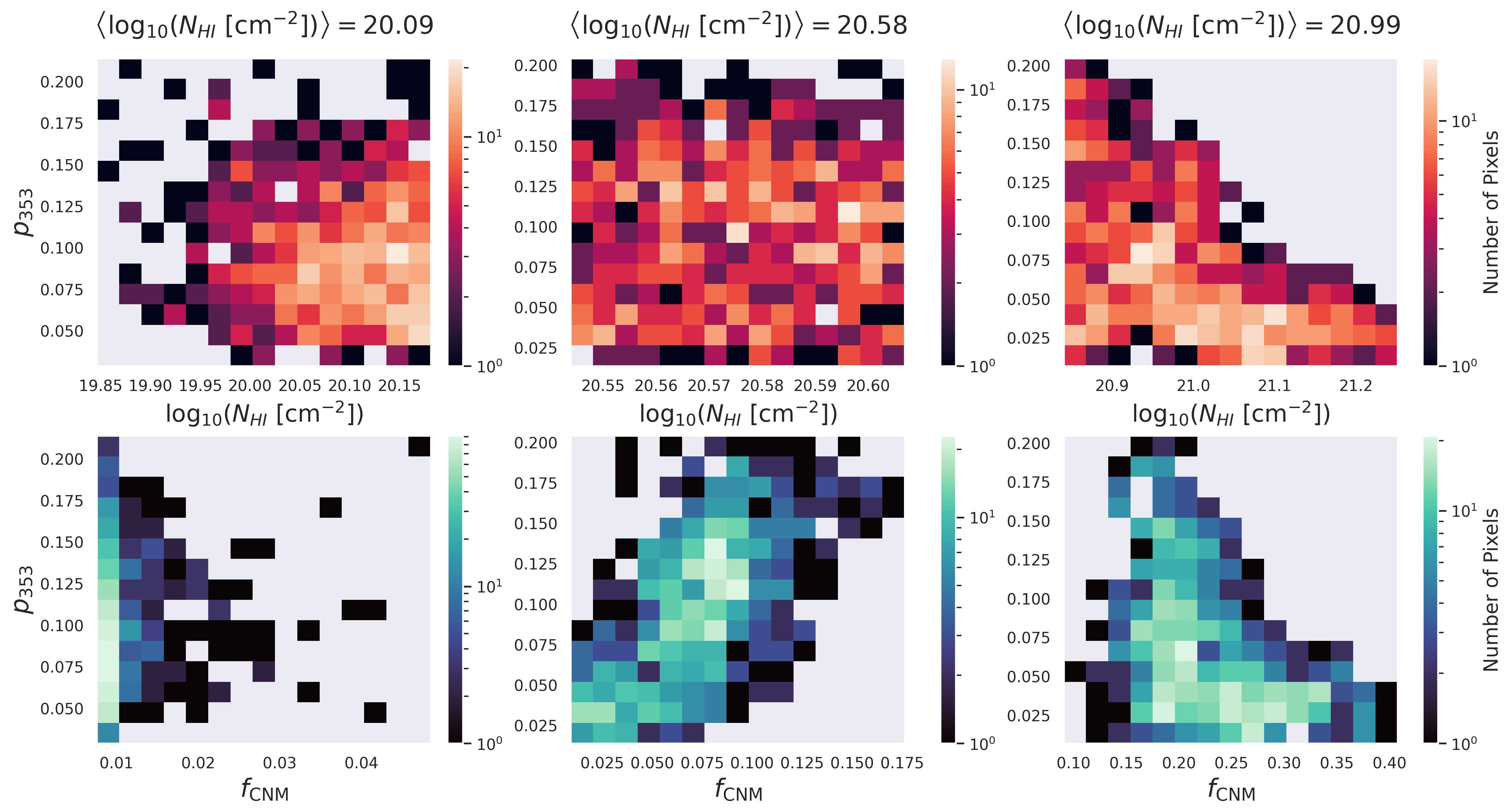}
\caption{Two-dimensional histograms of $p_{353}$ vs. $N_{\mathrm{H\,I}}$ (top row) and $p_{353}$ vs. $f_{\mathrm{CNM}}$ (bottom row) in low, intermediate, and high column density regions. At low column density ($\left<\log_{10}(N_{\mathrm{H\,I}} [\mathrm{cm}^{-2}])\right> < 20.09$), $\sim$ 99\% of sightlines have $f_{\mathrm{CNM}} < 0.03$ and $f_{\rm CNM}$ SNR $<$ 2. The limited $f_{\mathrm{CNM}}$ dynamic range and low $f_{\mathrm{CNM}}$ SNR makes the correlation result in this regime less reliable. In the intermediate column density regime, there is a clear positive $p_{353}$-$f_{\mathrm{CNM}}$ correlation while no trend is observed between $p_{353}$ and $N_{\mathrm{H\,I}}$. Finally, both $f_{\mathrm{CNM}}$ and $N_{\mathrm{H\,I}}$ anticorrelate with polarization fraction in the high column density regime. }
\label{fig:compare_col_den}
\end{figure*}

To investigate the hypothesis that magnetic field structure probed by dust polarization varies with ISM phase distribution, we test for a correlation between dust polarization fraction and CNM fraction ($f_{\mathrm{CNM}}$) in excess of any correlation between dust polarization fraction and \ion{H}{1} column density ($N_{\mathrm{H\,I}}$). If the dust polarization has no dependence on the \ion{H}{1} phase distribution, we expect no correlation between these parameters, except to the extent that both have some correlation with the total $N_{\mathrm{H\,I}}$. Using the 80\arcmin, offset-corrected, noise-debiased polarization fraction map $p_{353}$ described in Section \ref{subsec:pfrac}, we compute the correlation of $p_{353}$ with the $f_{\mathrm{CNM}}$ and $N_{\mathrm{H\,I}}$ maps smoothed to the same resolution in the high Galactic latitude ($|b|>30\degree$) sky. Motivated by the column-density-dependent $p_{353}$-$N_{\rm H}$ relation in Planck studies \citep{planck14-pd, planck14-mh, planck18-cr}, we calculate the Spearman correlation of $p_{353}$-$f_{\mathrm{CNM}}$ and $p_{353}$-$N_{\mathrm{H\,I}}$ in column density bins where each bin contains an equal number of independent measurements. 

\subsection{$p_{353}$-$N_\mathrm{H\,I}$ Correlation} \label{subsec:p353_nhi_corr}
The left panel of Figure \ref{fig:p353_fcnm_corr} shows the correlation between $p_{353}$ and the GALFA-\ion{H}{1} $f_{\mathrm{CNM}}$ and $N_{\mathrm{H\,I}}$ maps computed in ten column density bins. These bins correspond to mean $N_{\mathrm{H\,I}}$ values ranging from $10^{20}\ \mathrm{cm^{-2}}$ to $10^{21}\ \mathrm{cm^{-2}}$. The error bars on the correlation coefficients are estimated using a bootstrapping procedure, resampling the datasets with replacement for 1000 trials. The $p_{353}$-$N_{\mathrm{H\,I}}$ correlation result is consistent with prior results showing no correlation at low column densities and a strong anticorrelation in the highest $N_{\mathrm{H\,I}}$ bin (mean $\left<N_{\mathrm{H\,I}}\right>\sim10^{21}\ \mathrm{cm^{-2}}$) \citep{planck14-pd, planck18-cr}. In the top panels of Figure \ref{fig:compare_col_den}, we show the 2D histogram distribution of $p_{353}$ vs. $N_{\mathrm{H\,I}}$ in select column density regimes. Consistent with the Planck studies \citep{planck14-pd, planck18-cr}, $p_{353}$ shows significant scatter in low column-density regions and reaches $p_\mathrm{max}\sim0.2$. In the high column density regime, the range and mean $p_{353}$ decrease with increasing $N_{\mathrm{H\,I}}$, with $\left<p_{353}\right>\sim0.04$ at $N_{\mathrm{H\,I}}>10^{21}\ \mathrm{cm^{-2}}$.

\begin{figure*}[t]
\centering
\includegraphics[width=0.95\textwidth]{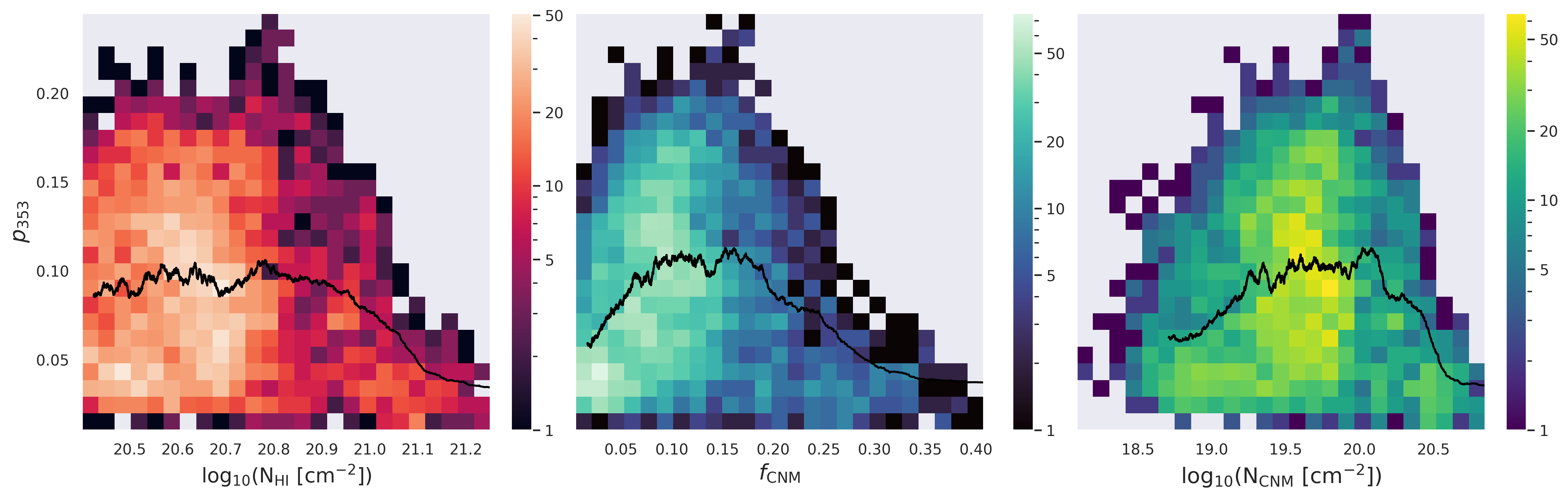}
\caption{Two-dimensional histograms of $p_{353}$-$N_{\mathrm{H\,I}}$ (left), $p_{353}$-$f_{\mathrm{CNM}}$ (middle), and $N_{\rm CNM}=f_{\mathrm{CNM}}N_{\mathrm{H\,I}}$ (right) in column density regime $\log_{10}(N_{\mathrm{H\,I}}\  [\mathrm{cm}^{-2}])>20.40$. The black curve overlaid on the histograms is the moving average of $p_{353}$ computed using a sliding window of size 100 sightlines. The average $p_{353}$ shows little variation at low $N_{\mathrm{H\,I}}$, but increases with increasing $f_{\mathrm{CNM}}$ or $N_{\rm CNM}$. All figures show a decreasing trend at high $N_{\mathrm{H\,I}}$, $f_{\mathrm{CNM}}$, and $N_{\rm CNM}$ values.}
\label{fig:p353_fcnm_hist2d}
\end{figure*}

\subsection{$p_{353}$-$f_{\mathrm{CNM}}$ Correlation} \label{subsec:p353_fcnm_corr}
In contrast with the $p_{353}$-$N_{\mathrm{H\,I}}$ relationship, we observe distinctly different behavior for the correlation between $p_{353}$ and $f_{\mathrm{CNM}}$. This is evident in the column density bins with $\left<\log_{10}\left(N_{\mathrm{H\,I}}~[\mathrm{cm^{-2}}]\right)\right><20.8$. As Figure~\ref{fig:p353_fcnm_corr} shows, in these regions where no $p_{353}$-$N_{\mathrm{H\,I}}$ relation is observed, there is a strong positive $p_{353}$-$f_{\mathrm{CNM}}$ correlation, with Spearman correlation coefficients up to $\sim0.6$. At higher column density regimes however, the degree of correlation decreases, and becomes consistent with the strong $p_{353}$-$N_{\mathrm{H\,I}}$ anticorrelation at the highest column density bin with $\left<N_{\mathrm{H\,I}}\right>\sim10^{21}\ \mathrm{cm^{-2}}$. To more clearly demonstrate this positive-to-negative correlation transition with increasing $N_{\mathrm{H\,I}}$, we show a map of the $p_{353}$-$f_{\mathrm{CNM}}$ correlation in the high Galactic latitude ($|b|>30\degree$) GALFA-\ion{H}{1} footprint in the righthand panel of Figure \ref{fig:p353_fcnm_corr}. To create the map, we divide the sky into 20 equal-area regions binned by column density, and color each region by its $p_{353}$-$f_{\mathrm{CNM}}$ correlation. Most regions in this portion of the high Galactic latitude sky show a strong positive correlation, except for a small, lower-Galactic-latitude ($-45\degree<b<-30\degree$) region of strong anticorrelation in the southern hemisphere. The adjacent regions show intermediate correlation values between that of the strongly-correlated low-column density regions and the anti-correlated high-column density regions. Overall, the $p_{353}$-$f_{\mathrm{CNM}}$ correlation map shows consistent behavior across large portions of the sky. The fact that the left and right panels of Figure \ref{fig:p353_fcnm_corr} show similar $p_{353}$-$f_{\mathrm{CNM}}$ relations despite dividing the sky into different numbers of equal-area regions further demonstrates that the correlation results are robust with respect to the particular choice of $N_{\mathrm{H\,I}}$ bins.

In Figure \ref{fig:compare_col_den} we examine the 2D histograms of $p_{353}$ vs. $f_{\mathrm{CNM}}$ and $p_{353}$ vs. $N_{\mathrm{H\,I}}$ for select low, intermediate and high column density regions selected from bins in {Figure \ref{fig:p353_fcnm_corr}}. At high column density with $\left<N_{\mathrm{H\,I}}\right>\sim10^{21}\ \mathrm{cm^{-2}}$, the $f_{\mathrm{CNM}}$ values range from $0.10\sim0.40$, while $p_{353}$ decreases from a maximum value of $\sim0.2$ to less than $0.04$ at peak $f_{\mathrm{CNM}}$. This is similar to the trend observed with increasing $N_{\mathrm{H\,I}}$. In the bin with intermediate column density values $\left<\log_{10}N_{\mathrm{H\,I}} [\mathrm{cm^{-2}}]\right>\sim20.6$, there is a striking linear correlation relation between $p_{353}$ and $f_{\mathrm{CNM}}$, with $p_{353}$ ranging from $\sim0$ to the maximum observed value of 0.2, while $f_{\mathrm{CNM}}$ ranges from approximately 0 to 0.17. No such correlation is observed between $p_{353}$ and $N_{\mathrm{H\,I}}$. The behaviors of the selected intermediate and high column density bins are representative of the full regions that show strong positive vs. negative $p_{353}$-$N_{\mathrm{H\,I}}$ correlations in Figure \ref{fig:p353_fcnm_corr}.

In the high Galactic latitude sky region we are considering, a significant majority of sightlines have low CNM content, with $f_{\mathrm{CNM}}<0.2$ \citep{murray18-sp, murray20-cn}. In fact, 90\% of the $|b|>30\degree$ GALFA-\ion{H}{1} footprint has $f_{\mathrm{CNM}}<0.17$. Thus, the ranges of the $p_{353}$ and $f_{\mathrm{CNM}}$ values in the regime where we observe a strong positive correlation are representative of the full map considered. This is not the case for the lowest column density bin with $\left<\log_{10}N_{\mathrm{H\,I}} [\mathrm{cm^{-2}}]\right>\sim20.1$, shown in the lower left plot of Figure \ref{fig:compare_col_den}. In this regime, the CNM content is extremely low, and low SNR, with over 99\% of the sightlines having $f_{\mathrm{CNM}}<0.03$ and $f_{\rm CNM}/\sigma_{f_{\rm CNM}}<2$. {We suspect that the loss of $p_{353}-f_{CNM}$ correlation here is driven by this loss of dynamic range in $f_{CNM}$: when most of the sightlines have $f_{\rm CNM}\sim0$, the $f_{\rm CNM}$-driven variation in $p_{353}$ is likely negligible compared to the general scatter in $p_{353}$, diluting any trend of $p_{353}$-$f_{\rm CNM}$ correlation we might otherwise observe over this lowest-column density region of the sky. }

\begin{figure*}[t]
\centering
\includegraphics[width=0.95\textwidth]{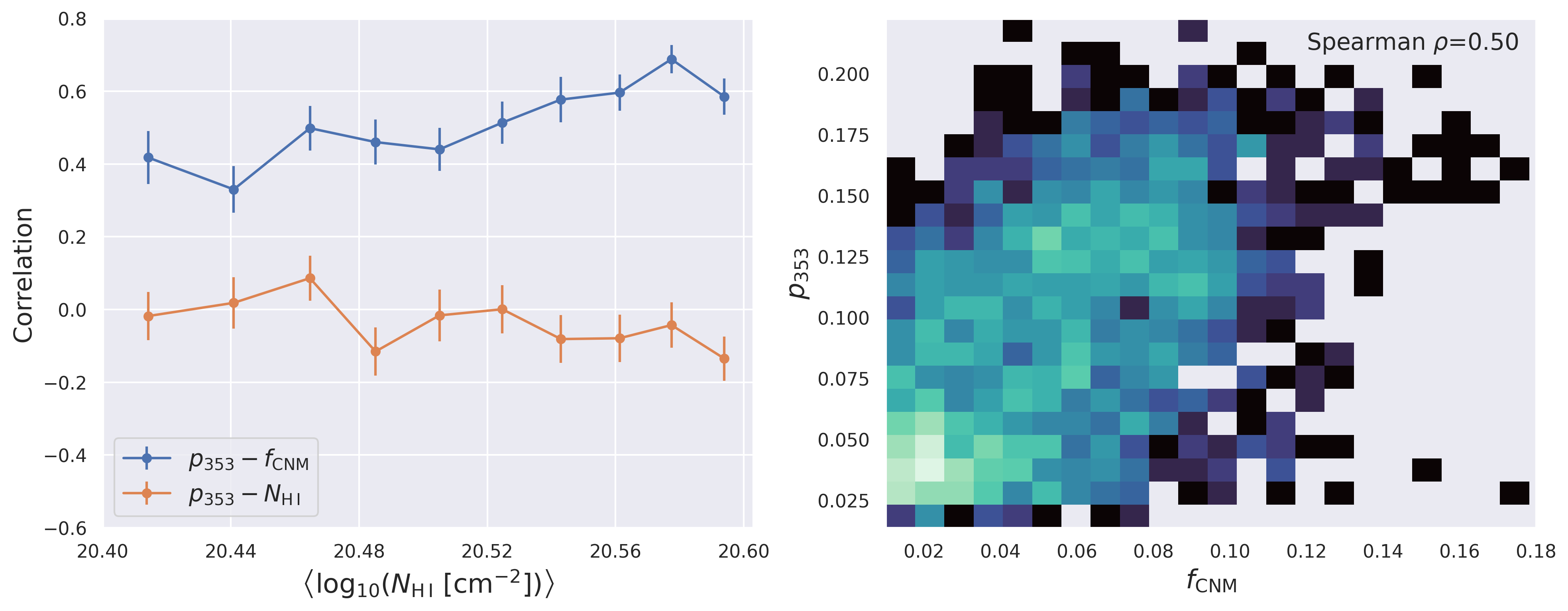}
\caption{Left: Comparing $p_{353}$-$f_{\mathrm{CNM}}$ vs. $p_{353}$-$N_{\mathrm{H\,I}}$ correlations in column density bins over the range $20.40<\log_{10}(N_{\mathrm{H\,I}}\  [\mathrm{cm}^{-2}])<20.6$. Right: 2D histogram of $p_{353}$ vs. $f_{\mathrm{CNM}}$ in the same regime. The lower and upper column density limits are motivated by $f_{\mathrm{CNM}}$ dynamic range and dust emission - $N_{\mathrm{H\,I}}$ linear correspondence respectively, as discussed in Section \ref{subsec:p353_fcnm_corr}. A consistent positive $p_{353}$-$f_{\mathrm{CNM}}$ correlation is observed across this regime, with an overall Spearman correlation coefficient of 0.50.}
\label{fig:p353_fcnm_corr_select}
\end{figure*}

\subsection{Column Density Range of Interest} \label{subsec:col_den_range}
We restrict our subsequent analysis to the range $\log_{10}(N_{\mathrm{H\,I}}\  [\mathrm{cm}^{-2}])>20.40$, where the ranges of $f_{\mathrm{CNM}}$ and $p_{353}$ are representative of the full high Galactic latitude sky considered and the mean $f_{\mathrm{CNM}}$ SNR is high ($>3$). Figure \ref{fig:p353_fcnm_hist2d} shows the overall 2D histogram distribution and running mean of the $p_{353}$-$N_{\mathrm{H\,I}}$, $p_{353}$-$f_{\mathrm{CNM}}$ and $p_{353}$-$N_{\mathrm{CNM}}$ (where $N_{\mathrm{CNM}} \equiv f_{\mathrm{CNM}}N_{\mathrm{H\,I}}$) distributions after applying this column density cut. $p_{353}$-$N_{\mathrm{H\,I}}$ shows the expected behavior of nearly constant $\left<p_{353}\right>$ at low $N_{\mathrm{H\,I}}$ and decreasing $\left<p_{353}\right>$ at high $N_{\mathrm{H\,I}}$. In contrast, the moving mean $\left<p_{353}\right>$ varies with $f_{\mathrm{CNM}}$ and $N_{\mathrm{CNM}}$ in a consistent way, increasing up to $f_\mathrm{CNM}\sim0.1$ and $N_{\mathrm{CNM}}\sim3\times10^{19}\rm~cm^{-2}$, before transitioning to an approximately monotonic decrease at higher CNM fraction and CNM column densities. 

At higher column densities, more of the hydrogen is in the molecular phase, making $N_{\mathrm{H\,I}}$ a less reliable tracer of dust \citep{burstein78-hi, lenz17-ln, nguyen18-hd}. Restricting to a lower column density range allows us to examine the observed positive $p_{353}$-$f_{\mathrm{CNM}}$ correlation in a regime where dust and \ion{H}{1} are largely tracing the same volumes. Using \ion{H}{1}4PI data, \citet{lenz17-ln} found that at $N_\mathrm{HI}<4\times10^{20}\ \mathrm{cm^{-2}}$, the correlation between the \ion{H}{1} and dust column is well represented by a linear fit with variations of less than 10\%. Thus we adopt $\log_{10}(4\times10^{20}\ \mathrm{cm^{-2}})\approx20.60$ as the upper limit. 

In Figure \ref{fig:p353_fcnm_corr_select}, we show the $p_{353}$-$f_{\mathrm{CNM}}$ and $p_{353}$-$N_{\mathrm{H\,I}}$ correlations in the selected column density range $20.40<\log_{10}(N_{\mathrm{H\,I}}\  [\mathrm{cm}^{-2}])<20.60$. $p_{353}$-$f_{\mathrm{CNM}}$ shows a consistent positive correlation in this range, with an overall Spearman correlation coefficient of 0.50. The right panel of Figure \ref{fig:p353_fcnm_corr_select} shows the 2d histogram distribution of $p_{353}$ vs. $f_{\mathrm{CNM}}$. The limit of $p_{353}$ ranges from 0 to observed $p_{max}$ for the full map at $\sim0.2$, while $f_{\mathrm{CNM}}$ ranges from 0 to 0.18, approximately the 90th percentile of the full map $f_{\mathrm{CNM}}$ distribution. Thus the dynamic ranges of $p_{353}$ and $f_{\mathrm{CNM}}$ in this region are representative of the full high Galactic latitude sky. While $20.40<\log_{10}(N_{\mathrm{H\,I}}\  [\mathrm{cm}^{-2}])<20.60$ covers only a limited range of the full $N_{\mathrm{H\,I}}$ distribution, it accounts for $\sim27\%$ of the high Galactic latitude GALFA-\ion{H}{1} sky for a total of $1850~\rm deg^2$. As Figure \ref{fig:p353_fcnm_corr} shows, the positive $p_{353}$-$f_{\mathrm{CNM}}$ correlation extends beyond the column density range we select here, and it would be interesting to examine the transition from correlation to anticorrelation at higher column densities in future work.

In Appendix \ref{appx:galfa_v_hi4pi}, we repeat the same analysis with \ion{H}{1}4PI column density and $f_\mathrm{CNM}$ data, and {find the same qualitative result that there is a statistically significant $p_{353}$-$f_\mathrm{CNM}$ correlation in the regimes where no $p_{353}$-$N_{\mathrm{H\,I}}$ relation is observed. However, the degree of correlation over the \ion{H}{1}4PI sky is smaller compared to that in the GALFA-\ion{H}{1} footprint. We argue that a significant factor in that difference is the more limited $f_{\rm CNM}$ dynamic range and a higher proportion of low $f_{\rm CNM}$ SNR sightlines. The arguments are presented in more detail in Appendix \ref{appx:galfa_v_hi4pi}. Here, we focus our analysis on the GALFA-\ion{H}{1} sky data, which is a general footprint spanning $\sim40\%$ of the high latitude ($|b|>30\degree$) sky across different environments. }

In short, exploring the high-latitude diffuse ISM using GALFA-\ion{H}{1} and Planck dust maps, we found a strong correlation between polarization fraction $p_{353}$ and CNM mass fraction $f_\mathrm{CNM}$. The positive correlation is present in column density regimes where there is no $p_{353}$-$N_{\mathrm{H\,I}}$ correlation. 

\section{Interpreting The Dust Polarization - CNM Fraction Correlation} \label{sec:corr_interp}
In this section, before testing the hypothesis of relative disorder of magnetic fields between phases, we first examine other possible contributions to dust polarization fraction variation, and show that they cannot account for the observed positive $p_{353}$-$f_{\rm CNM}$ correlation. Following the formalism for thermal dust emission presented in \citet{fiege00-pe} \citep[see also][]{pelkonen06-pe}, for emission at sub-millimeter wavelengths assuming uniform grain properties, the Stokes $Q$ and $U$ can be parameterized as the following integral along the LOS:
\begin{align}
    Q &= \int \alpha \epsilon \rho \cos{2\psi}\cos^2{\gamma}\,ds \label{eq:q} \\
    U &= \int \alpha \epsilon \rho \sin{2\psi}\cos^2{\gamma}\,ds \label{eq:u}
\end{align}
where $\psi$ is the POS magnetic field angle, and $\gamma$ is the inclination angle of the magnetic field relative to the {POS}. $\rho$ is the gas density. {$\epsilon$ is the dust emissivity}. $\alpha$ is a grain property coefficient parameterizing grain alignment and polarizing efficiency. In practice, we determine $\alpha$ from the maximum intrinsic polarization fraction $p_{\rm max}$: 
\begin{equation} \label{eq:alpha}
    p_\mathrm{max} = \frac{\alpha}{1-\alpha/6}
\end{equation}
The polarization fraction is computed from the Stokes parameters as:
\begin{equation} \label{eq:pfrac}
    p = \frac{P}{I}=\frac{\sqrt{Q^2+U^2}}{I},
\end{equation}
where the intensity $I$ can be further expressed as
\begin{equation} \label{eq:int}
    I =\int\epsilon\rho\,ds-\frac{1}{2}\int\alpha\epsilon\rho(\cos^2{\gamma}-\frac{2}{3})\,ds.
\end{equation}

There are two main factors affecting the observed dust polarization fraction along a given sightline: grain alignment efficiency parameterized by $\alpha$, and magnetic field structures characterized by the magnetic field angles $\psi$ and $\gamma$. We examine how each factor might contribute to a positive $p_{353}$-$f_\mathrm{CNM}$ correlation.

\subsection{Correlation Inconsistent with Variable Grain Alignment} \label{subsec:grain_align}
Variation in grain alignment efficiency can affect dust polarization fraction. For example, in dense molecular regions, dust grains are shielded from the UV and optical radiation that is responsible for initiating the dust spinning and alignment process. This results in less efficient alignment with the local magnetic field and consequently lower polarization fraction \citep{draine96-sp, hoang08-sp, andersson15-re}. While loss of alignment could play a role in dense molecular cloud regions, in the diffuse region away from the Galactic plane selected in our study ($20.40<\log_{10}(N_{\mathrm{H\,I}}\  [\mathrm{cm}^{-2}])<20.60$), the column density and molecular content never get high enough for the shielding of dust grains to become a factor. Furthermore, CNM is the colder, denser phase of \ion{H}{1}. Even if there were some degree of alignment loss with increasing CNM content, it would result in an anticorrelation between $p_{353}$ and $f_{\mathrm{CNM}}$, the opposite of the observed positive $p_{353}$-$f_{\mathrm{CNM}}$ relation. {Variation in grain alignment efficiency can potentially account for some dispersion in the polarization fraction distribution. \citet{medan18-ga} found that a model of varying grain alignment due to the non-uniform distribution of the nearby OB associations fits well the spatial variation of starlight polarization data along Local Bubble wall regions. However, we do not expect such radiation field variations to correlate with CNM content and drive the $p-f_{\rm CNM}$ relation. Similarly, variation of other grain properties such as size, porosity, and composition might affect the general dispersion of $p$ \citep{draine21-ei}, but should not directly contribute to a $p_{353}$-$f_{\mathrm{CNM}}$ relation over a large region of the diffuse sky. } Thus, we can rule out grain alignment efficiency as a contributor to the $p_{353}$-$f_{\mathrm{CNM}}$ correlation.

\begin{figure*}[t]
\centering
\includegraphics[width=\textwidth]{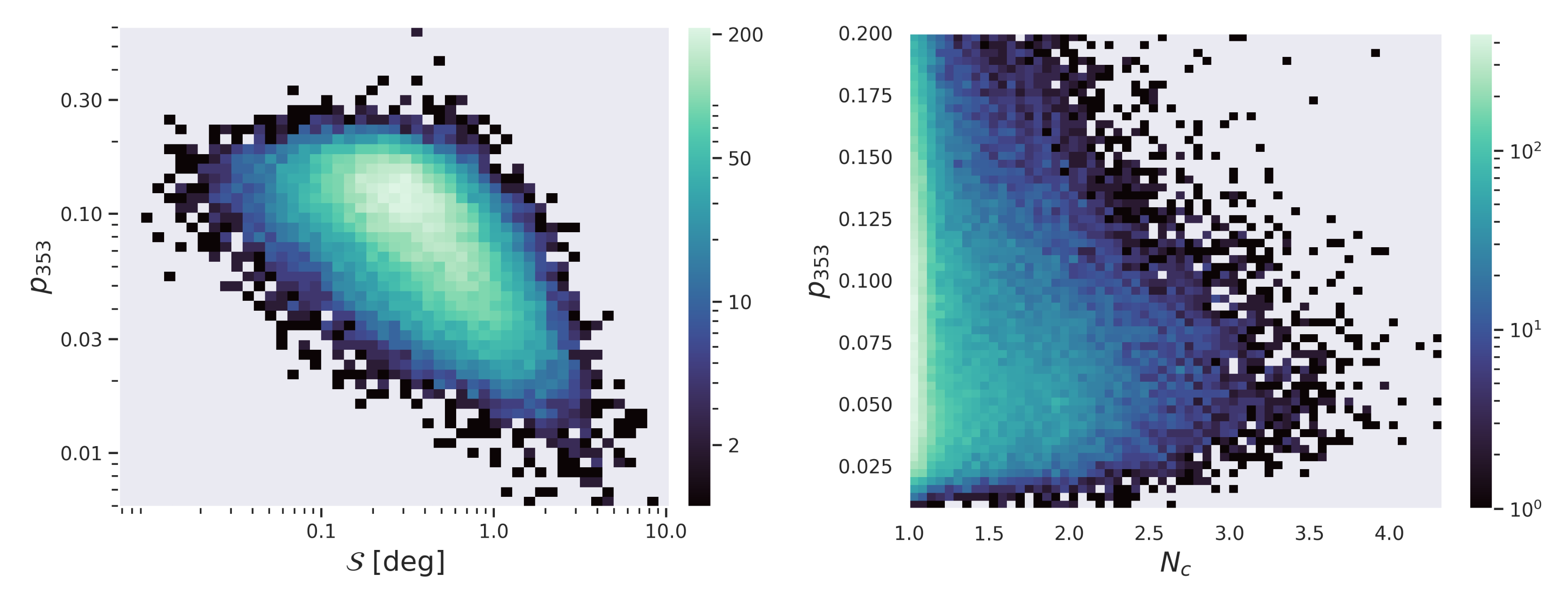}
\caption{Two-dimensional histograms of $p_{353}$ vs. $\mathcal{S}$ (left) and $p_{353}$ vs. $N_c$ (right) in the high Galactic latitude ($|b|>30\degree$) GALFA-\ion{H}{1} sky. Regions of high polarization angle dispersion generally correspond to lower polarization fraction, while regions of high LOS complexity correspond to a decrease in the range and the mean value of $p_{353}$. }
\label{fig:p353_s_nc_2d}
\end{figure*}

\begin{figure*}[t]
\centering
\includegraphics[width=\textwidth]{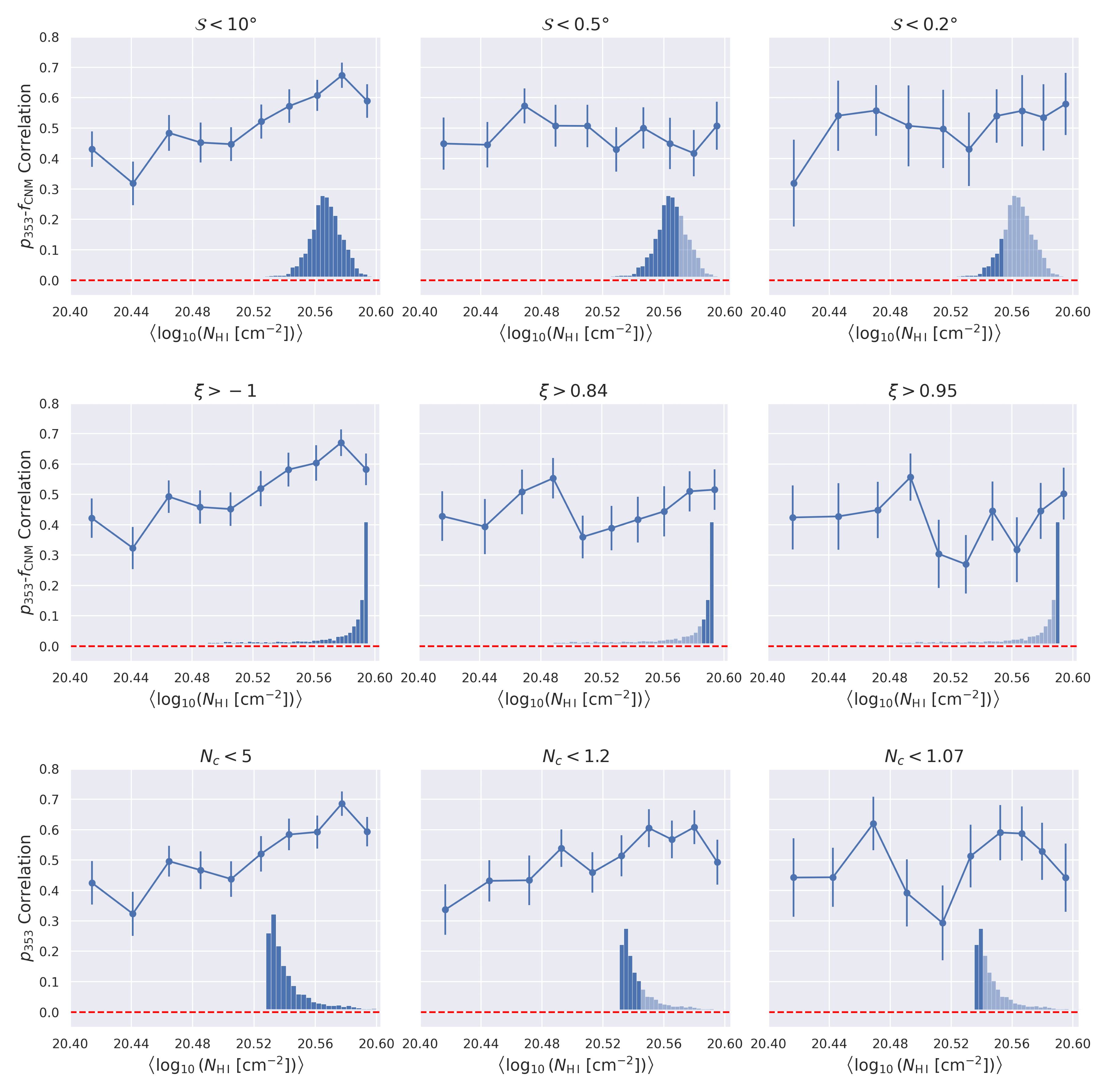}
\caption{Column-density-binned $p_{353}$-$f_{\mathrm{CNM}}$ correlation over masks defined by different values of the polarization angle dispersion $\mathcal{S}$ (top), degree of alignment between \ion{H}{1} and dust polarization angles $\xi$ (middle), and LOS complexity $N_c$ (bottom). The distribution of $\mathcal{S}$, $\xi$, $N_c$ are shown at the bottom left of each panel. From left to right, the limits are set to represent approximately 100\%--60\%--20\% of the data. The strong $p_{353}$-$f_{\mathrm{CNM}}$ correlation persists across different $\mathcal{S}$, $\xi$, $N_c$ value limits. }
\label{fig:p353_fcnm_lim}
\end{figure*}


\subsection{Correlation Not Driven By POS Dispersion} \label{subsec:incl_angle}
One aspect of magnetic field structure that affects polarization fraction is the variation of the inclination angle $\gamma$. As specified in Equations \ref{eq:q}--\ref{eq:pfrac}, $p$ is maximum when the magnetic field is entirely in the POS. In the general case where $\gamma$ is nonzero, the $p$ will be lower than its theoretical maximum, and is zero when the magnetic field is perfectly parallel to the LOS. Inclination angle is also related to the magnetic field angle dispersion in the POS. When the mean magnetic field is nearly parallel to the LOS, small perturbations to the 3D orientation of the magnetic field will result in large changes to its POS projection, resulting in depolarization due to disordered magnetic field orientations along the LOS and within the beam. 
However, in this study we do not expect these effects to translate into a significant contribution to the $p_\mathrm{353}$-$f_\mathrm{CNM}$ correlation, since we do not expect $\gamma$ to correlate with $f_\mathrm{CNM}$ across large regions of the diffuse sky that are not physically connected. We test that expectation by examining the variation of $p_\mathrm{353}$-$f_\mathrm{CNM}$ correlation in regions of different degrees of POS polarization angle dispersion.

Magnetic field dispersion in the POS can be characterized by the polarization angle dispersion function:
\begin{equation} \label{eq:disp_func}
    \mathcal{S}(\boldsymbol{r},\delta)=\sqrt{\frac{1}{N}\sum_{i=1}^{N}[\psi(\boldsymbol{r}+\boldsymbol{\delta_i})-\psi(\boldsymbol{r})]^2}
\end{equation}
where $i$ sums over $N$ pixels within an annulus of inner radius $\delta/2$ and outer radius $3\delta/2$. We estimate the dispersion function from the Stokes $Q$ and $U$ parameters, at 80\arcmin\ resolution with a lag $\delta$=40\arcmin\ \citep{planck18-cr}. $\mathcal{S}$ quantifies the local variance of the polarization angle on the POS. In the left panel of Figure \ref{fig:p353_s_nc_2d}, we show the distribution of $p_{353}$ vs. $\mathcal{S}$ in the high-latitude GALFA-\ion{H}{1} footprint. Consistent with the results from \citet{planck18-cr}, regions of high polarization angle dispersion correspond to lower polarization fraction. 

We also make use of another quantity related to polarization angle dispersion discussed in \citet{clark19-hs}. When comparing the \ion{H}{1}-based and dust 353 GHz polarization maps, the authors also computed the angular difference between \ion{H}{1} and dust polarization angles:
\begin{equation} \label{eq:dtheta}
    \delta\theta=\frac{1}{2}\arctan{\left[\frac{\sin{2\theta_\mathrm{H\,I}}\cos{\theta_{353}}-\cos{2\theta_\mathrm{H\,I}}\sin{\theta_{353}}}{\cos{2\theta_\mathrm{H\,I}}\cos{\theta_{353}}+\sin{2\theta_\mathrm{H\,I}}\sin{\theta_{353}}}\right]}
\end{equation}
where $\theta_\mathrm{H\,I}$ and $\theta_{353}$ are the \ion{H}{1} orientation and dust 353 GHz polarization angles respectively. They further define from $\delta\theta$ a measure of the mean degree of alignment:
\begin{equation} \label{eq:xi}
    \xi=\left<\cos{2\delta\theta}\right>
\end{equation}
so that $\xi=1$ corresponds to perfect alignment, and $\xi=-1$ corresponds to anti-alignment where $\theta_\mathrm{H\,I}$ and $\theta_{353}$ are perpendicular to each other. Examining the degree of alignment $\xi$ as a function of \ion{H}{1}-based dispersion function $\mathcal{S}$, they found a strong anti-correlation, suggesting that regions of low dispersion where the mean magnetic field is more likely to be in the POS, are also where the measured \ion{H}{1} and dust polarization angles are preferentially aligned.

To empirically test whether these effects drive the observed $p_\mathrm{353}$-$f_\mathrm{CNM}$ relation, we examine if there is a loss of correlation in regions of increasingly narrow $\mathcal{S}$ and $\xi$ ranges. The results are shown in the top and middle panels of Figure \ref{fig:p353_fcnm_lim}. The specific limits of $\mathcal{S}$ and $\xi$ are determined so that the panels from left to right account for approximately 100\%--60\%--20\% of the data respectively. The degree of $p_\mathrm{353}$-$f_\mathrm{CNM}$ correlation is consistent even at low $\mathcal{S}$ and high $\xi$ limits, suggesting that the correlation trend persists when the magnetic field mostly lies on the plane of the sky, and POS dispersion does not drive the $p_{353}$-$f_{\rm CNM}$ correlation. 

{There are regions in the high latitude sky where a phase-related $\gamma$ variation might be expected. \citet{skalidis19-lb} showed that in Galactic polar cap ($|b|>60\degree$) regions, the Planck 353 GHz polarized emission is dominated by close-by magnetized structures that coincide with the Local Bubble wall. Modeling magnetic field structure in the same region, \citet{pelgrims20-lb} found that the mean magnetic field in each polar cap is closely aligned with the POS. If the CNM at high Galactic latitude is dominated by closer-by structures associated with the Local Bubble while the WNM extends much farther out, then there could be a phase-dependent $\gamma$ variation in those regions. However, most of the sightlines ($\sim 87\%$) in our region of interest lie at lower Galactic latitudes than the polar cap regions examined in these works. For $\gamma$ variation to drive the $p_{353}-f_{\rm CNM}$ correlation, there must be a clear offset in the mean $\gamma$ angle between the CNM and WNM. While some WNM structures might extend much farther out in distance, there is no evidence for a clear separation in physical location between the CNM structures and the bulk of the WNM column. Moreover, analyzing the polarization fraction from the same Planck 353 GHz dataset as our study in the high latitude regions, \citet{halal24-lb} find that there is no clear association between $p_{353}$ and the tangent plane of the Local Bubble wall. Instead, those authors found that $p_{353}$ variation is more correlated with the complexity of dust structures along the LOS. This implies that the measured dust polarization is in general sensitive to extended structures both within and beyond the Local Bubble wall. Thus we do not expect a significant phase-dependent $\gamma$ variation due to association with the Local Bubble to drive the observed $p_{353}$-$f_{\rm CNM}$ relation. }

\subsection{Correlation Not Driven by LOS \ion{H}{1} Components} \label{subsec:los_depol}

The presence of multiple dust clouds -- traced by multiple \ion{H}{1} components -- along the LOS can also contribute to the polarization fraction variation. The basic picture is that sightlines dominated by a single \ion{H}{1} cloud are more likely to represent a single dust cloud with a coherent ambient mean magnetic field than sightlines with multiple \ion{H}{1} clouds, which may in general be separated in distance along the LOS \citep{pelgrims21-ld}. Here we examine whether this effect can account for the observed $p_{353}$-$f_\mathrm{CNM}$ relation without explicit assumptions about how the magnetic field structure varies across phases. We use the column-density weighted LOS complexity measure $N_c$ discussed in Section \ref{subsec:ncloud} and defined in Equation~\ref{eq:nc}. We show the 2D histogram of $N_c$-$p_{353}$ in the right panel of Figure~\ref{fig:p353_s_nc_2d}. We observe a decrease of the range and mean $p_{353}$ values with increasing LOS complexity $N_c$, consistent with what \citet{panopoulou20-nc} found.

To test if the $p_{353}$-$N_c$ relation might drive the observed $p_{353}$-$f_\mathrm{CNM}$ correlation, we examine whether there is a loss of $p_\mathrm{353}$-$f_\mathrm{CNM}$ correlation when we restrict to narrower ranges of $N_c$. As the bottom panels of Figure \ref{fig:p353_fcnm_lim} show, similar to the $\mathcal{S}$ and $\xi$ results in Section \ref{subsec:incl_angle}, we find that the strong positive $p_\mathrm{353}$-$f_\mathrm{CNM}$ correlation persists even when limited to a narrow range of $N_c$ values, and at low $N_c\sim1$ where the \ion{H}{1} emission is dominated by a single \ion{H}{1} cloud.

In conclusion, we argued that grain property variation does not play a significant role in the high-latitude diffuse sky, and showed that the relationships between $p_{353}$ and magnetic field properties like inclination angle $\gamma$, POS dispersion $\mathcal{S}$, and LOS complexity $N_c$ are not sufficient to produce the observed positive $p_\mathrm{353}$-$f_\mathrm{CNM}$ correlation. Thus, the correct explanation most likely requires an explicit assumption on the relative disorder of the magnetic field between interstellar volumes occupied by different \ion{H}{1} phases. We examine this hypothesis in the next section. 

\begin{figure}[t]
\centering
\includegraphics[width=0.5\textwidth]{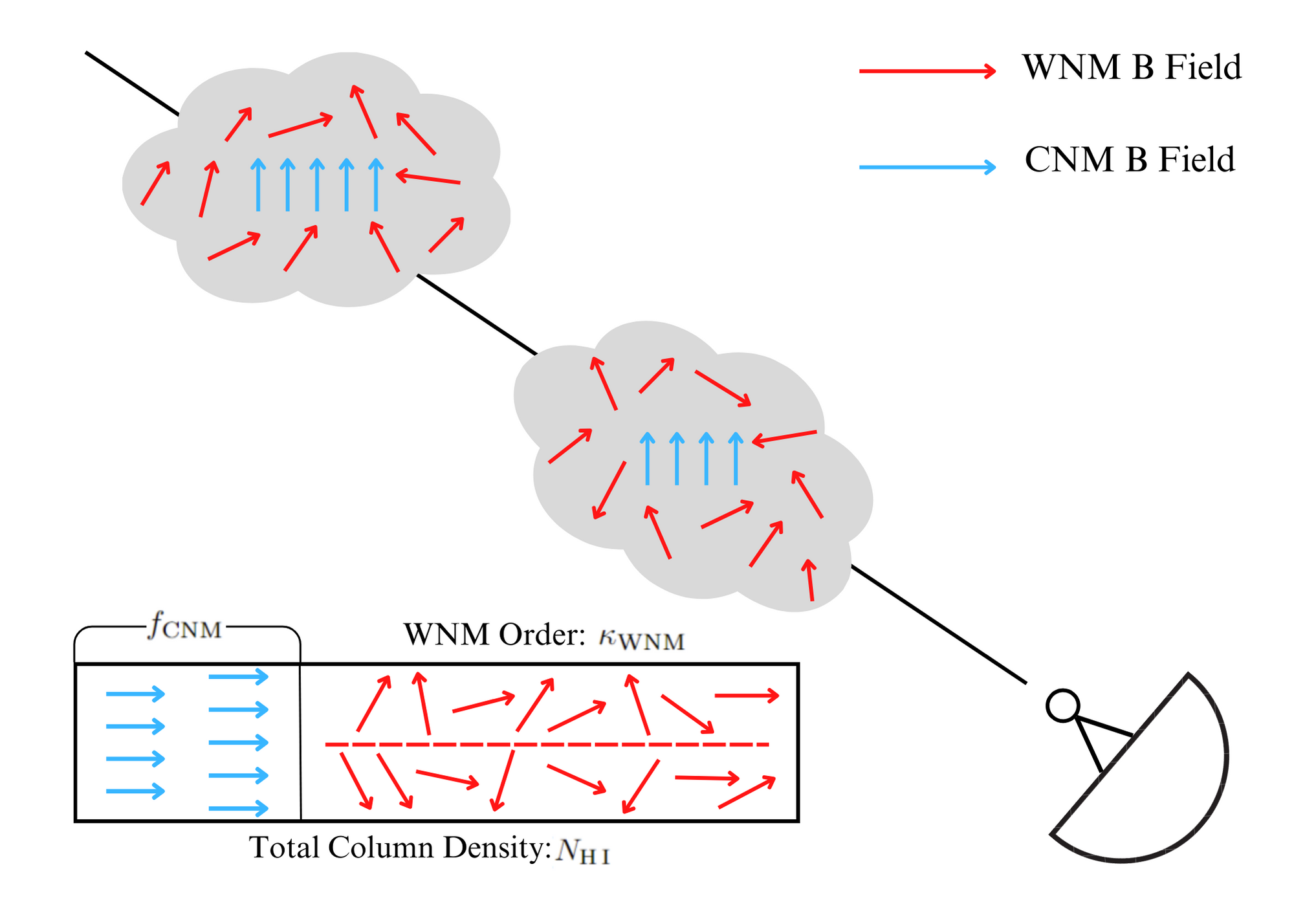}
\caption{Cartoon illustration of relative disorder of magnetic fields between CNM and WNM as observed along a given line of sight described by model1. If the WNM magnetic field is more disordered than that of the CNM as illustrated in the figure, then sightlines with higher CNM fraction will suffer from less LOS depolarization, resulting in a positive $p_{353}$-$f_{\mathrm{CNM}}$ correlation. In this cartoon, the relative fraction of the column occupied by each phase (bottom rectangle) is representative of the real sky, but the relative fraction of the ISM volume (cartoon clouds) is not.}
\label{fig:cartoon_coherence}
\end{figure}


\section{Modeling the Dust Polarization - \ion{H}{1} Phase Connection} \label{sec:corr_model}
Having ruled out other factors discussed in Section \ref{sec:corr_interp} as drivers of the observed $p_{353}$-$f_{\rm CNM}$ relation, here we argue that the correlation is best reproduced by explicitly modeling LOS magnetic field disorder as dependent on \ion{H}{1} phase. This disorder is characterized by the variation of the POS angle $\psi$ in Equations \ref{eq:q} and \ref{eq:u}, where a more uniform magnetic field adds constructively when integrated along the LOS, resulting in higher observed $p$. Thus, if there is a difference in the magnetic field disorder between different phases of the ISM, a relationship between $p$ and phase content, like $p_{353}$-$f_{\mathrm{CNM}}$ naturally follows. Specifically, the positive $p_{353}$-$f_{\mathrm{CNM}}$ correlation discussed here could point to a more magnetically ordered CNM compared to more tangled magnetic field structures in the WNM. Because the CNM occupies a much smaller volume than the WNM, a relatively ordered magnetic field structure on CNM length scales is a reasonable starting assumption. In other words, any region of WNM will occupy a much larger path length than a region of equivalent-column-density CNM, and thus will in general experience more variation in the magnetic field orientation. Here, we quantitatively constrain the additional LOS depolarization that is attributable to the WNM-associated dust. 
First, we present a series of cartoon models that parameterize the separate contributions to the integrated dust polarization signal from the dust associated with the WNM gas and that associated with the CNM gas. For simplicity we will refer to the magnetic fields in these regions as the ``CNM magnetic field" and ``WNM magnetic field" from now on. Since the $f_\mathrm{CNM}$ data we utilize here measures the CNM component as a fraction of the total \ion{H}{1} column, here we use WNM to denote anything not captured by the $f_\mathrm{CNM}$ measurement, which in general also includes a contribution from the thermally unstable medium. {The total dust column in principle also includes any dust associated with the warm ionized medium or undetected molecular gas, although we expect these contributions to be small.} 

\subsection{Phase-Dependent Cartoon Models} \label{subsec:cartoon_model}
To explicitly model the dependence of the polarization fraction on the variation of the POS component of magnetic fields along the LOS across \ion{H}{1} phases, we follow the formalism presented in Equations \ref{eq:q}--\ref{eq:int}. Based on the result of Section \ref{subsec:incl_angle} that $\gamma$ variation does not contribute meaningfully to the $p_\mathrm{353}$-$f_\mathrm{CNM}$ correlation, we assume without loss of generality that $\gamma=0$. To model varying magnetic field disorder across \ion{H}{1} phases, we rewrite Equations \ref{eq:q} and \ref{eq:u} into contributions from separate WNM and CNM components. For simplicity, we further assume uniform volume densities in each phase ($\rho_{\rm CNM}$, $\rho_{\rm WNM}$) to rewrite the equations in terms of CNM and WNM column densities. We denote all model quantities in this setup ``model1" to distinguish it from an updated model we will introduce later:
\begin{align}
    Q_{\rm model1} &= \sum_i^\mathrm{phase} \sum_s^{\rm LOS} \alpha \epsilon N_i \cos{2\psi_i(s)} \label{eq:qm1} \\
    U_{\rm model1} &= \sum_i^\mathrm{phase} \sum_s^{\rm LOS} \alpha \epsilon N_i \sin{2\psi_i(s)} \label{eq:um1}
\end{align}
where $i=$ CNM or WNM, and $N_{\rm CNM(WNM)}=f_{\rm CNM(WNM)}N_{\rm H\,I}=\int\rho_{\rm CNM(WNM)}\,ds$ is the CNM (WNM) column density. {$\alpha$ is a constant related to $p_{\rm max}$ as specified by Equation \ref{eq:alpha}}. {$\epsilon$ is the dust emissivity. In the diffuse, low-density ($N_\mathrm{H\,I}<4\times10^{20}~\mathrm{cm^{-2}}$) region considered in this study we do not expect significant emissivity variation \citep{lenz17-ln}. We will discuss the effect of $\epsilon$ variability further when discussing the interpretation of the best-fit parameters. Here for simplicity, we assume uniform $\epsilon$ which cancels out when computing the model polarization fraction $p_{model1}$ from $Q_{\rm model1}$, $U_{\rm model1}$ and the total column density $N_{\rm HI}$}:
\begin{equation} \label{eq:pm1}
    p_{\rm model1} = \sqrt{Q_{\rm model1}^2+U_{\rm model1}^2} / \sum_i^\mathrm{phase}{N_i}
\end{equation}
The magnetic field disorder of a given phase is encoded by the variation of $\psi_i(s)$ along the LOS. {However, regardless of how $\psi_i(s)$ is distributed along the LOS, what is directly constrained by the relationship between $p_{353}$ and $f_{\rm CNM}$ is the depolarization in the CNM column vs. the WNM column. Thus, we can reduce the LOS summation over $\psi_i(s)$ in each phase to two geometric depolarization factors $\kappa_{\rm CNM}$ and $\kappa_{\rm WNM}$. A sightline with $f_{\rm CNM}=0$ will have polarization fraction $p_{\rm WNM} = \kappa_{\rm WNM}p_{\rm max}$, while a $f_{\rm CNM}=1$ sightline has $p_{\rm CNM} = \kappa_{\rm CNM}p_{\rm max}$. The polarization fraction of a general sightline will then be determined by $p_{\rm CNM}$ and $p_{\rm WNM}$ weighted by the respective phase fractions $f_{\rm CNM}$ and $(1- f_{\rm CNM})$, as well as the angle between the mean CNM and WNM polarization angle orientations.} 

{We make a further simplification for our cartoon model by assuming that the mean polarization angles in the CNM and WNM columns are aligned. This is motivated by the strong correlation between dust polarization and \ion{H}{1} polarization templates constructed from the orientation of primarily CNM \ion{H}{1} filaments \citep[][see Section \ref{subsec:hi_pol_int}]{clark19-hs}. Because the \ion{H}{1} templates over-weight the contribution of the CNM column to the polarized emission, the agreement between dust polarization and the \ion{H}{1} filament orientation constrains how much the mean magnetic field orientation in the WNM column can differ from that in the CNM column, especially in diffuse regions where the filamentary CNM structures account for a minority of the column density \citep{kalberla16-hi}. The median degree of alignment $\xi$ between $\theta_{353}$ and the \ion{H}{1}-based angle $\theta_{H\,I}$ as defined by Equations \ref{eq:dtheta} and \ref{eq:xi}, is $\sim0.92$ in our region of interest, where $\xi=1$ indicates perfect alignment. Assuming $\theta_{H\,I}$ traces the CNM orientation while $\theta_{353}$ is a weighted vector sum of both the CNM and WNM orientation, we can derive the maximum difference between the CNM and WNM mean angles that could still lead to a $\xi_{H\,I-353}\sim0.92$ degree of alignment. For each sightline in our dataset, we assign a random orientation $\theta_{CNM}$ to the CNM component. The WNM orientation is then given by $\theta_{WNM}=\theta_{CNM}+\Delta \theta$, where $\Delta \theta$ is the mean angle offset between the WNM and the CNM. Assuming $\theta_{H\,I}=\theta_{CNM}$, and $\theta_{353}$ is the weighted vector sum of $\theta_{CNM}$ and $\theta_{WNM}$, we can compute a mean $\xi$ across sightlines and constrain the maximum $\Delta \theta_{\rm max}$. We assign a weight of $C N_{\rm CNM}$ to the CNM angles and $N_{\rm WNM}$ to the WNM angles. The column density weightings are derived from the $f_{\rm CNM}$ and $N_{H\,I}$ data for each sightline, while $C$ accounts for any additional weighting due to factors such as a higher emissivity or lower degree of internal depolarization in the CNM column. When $C=1$, we find that a median alignment of $\xi=0.92$ limits $\Delta \theta$ to be at maximum $\sim11\degree$. Even a significantly higher CNM weighting at $C=10$ only changes the result by less than $5\degree$. Thus it is reasonable to assume in our region of interest that the mean CNM and WNM magnetic field orientations are aligned and reduce the vector sum to a scalar sum. The equation for $p_{\rm model1}$ can then be rewritten as}
\begin{align} \label{eq:pm1_fcnm}
    p_{\rm model1} &= f_{\rm CNM}\cdot p_{\rm CNM} + (1- f_{\rm CNM})\cdot p_{\rm WNM} \\ \nonumber
                   &= p_{\rm max}\bigl(f_{\rm CNM}\kappa_{\rm CNM}+(1- f_{\rm CNM})\kappa_{\rm WNM}\bigr) \\ \nonumber
                   &= p_{\rm max}(\kappa_{\rm CNM}-\kappa_{\rm WNM})f_{\rm CNM}+p_{\rm max}\kappa_{\rm WNM}
\end{align}
{This is a linear relation between $p_{\rm model1}$ and $f_{\rm CNM}$ with slope $p_{\rm max}(\kappa_{\rm CNM}-\kappa_{\rm WNM})$ and intercept $p_{\rm max}\kappa_{\rm WNM}$ that we can fit to the observed $p_{353}-f_{\rm CNM}$ correlation. We further define $\kappa=\kappa_{\rm WNM}/\kappa_{\rm CNM}$ to encode the relative depolarization in the WNM vs. CNM columns. The (slope, intercept) pair then becomes ($p_{\rm max}\kappa_{\rm CNM}(1-\kappa)$), $p_{\rm max}\kappa_{\rm CNM}\kappa$). Written this way, $\kappa_{\rm CNM}$ is entirely degenerate with $p_{\rm max}$ in a linear fit. So without loss of generality, we simplify our parameterization by setting $\kappa_{\rm CNM}=1$. The linear relation between $p_{\rm model1}$ and $f_{\rm CNM}$ simplifies to:}
\begin{align} \label{eq:plinear}
    p_{\rm model1} &= p_{\rm max}(1-\kappa)f_{\rm CNM}+p_{\rm max}\kappa
\end{align}

Thus, fitting this model to the observed $p_{353}$-$f_{\rm CNM}$ data is equivalent to deriving a best-fit linear relation, where the slope is parameterized by $p_{\rm max}(1-\kappa)$, and the intercept by $p_{\rm max}\kappa$. We illustrate the setup of this simple cartoon model in Figure \ref{fig:cartoon_coherence}.

\begin{figure}[t]
\centering
\includegraphics[width=0.45\textwidth]{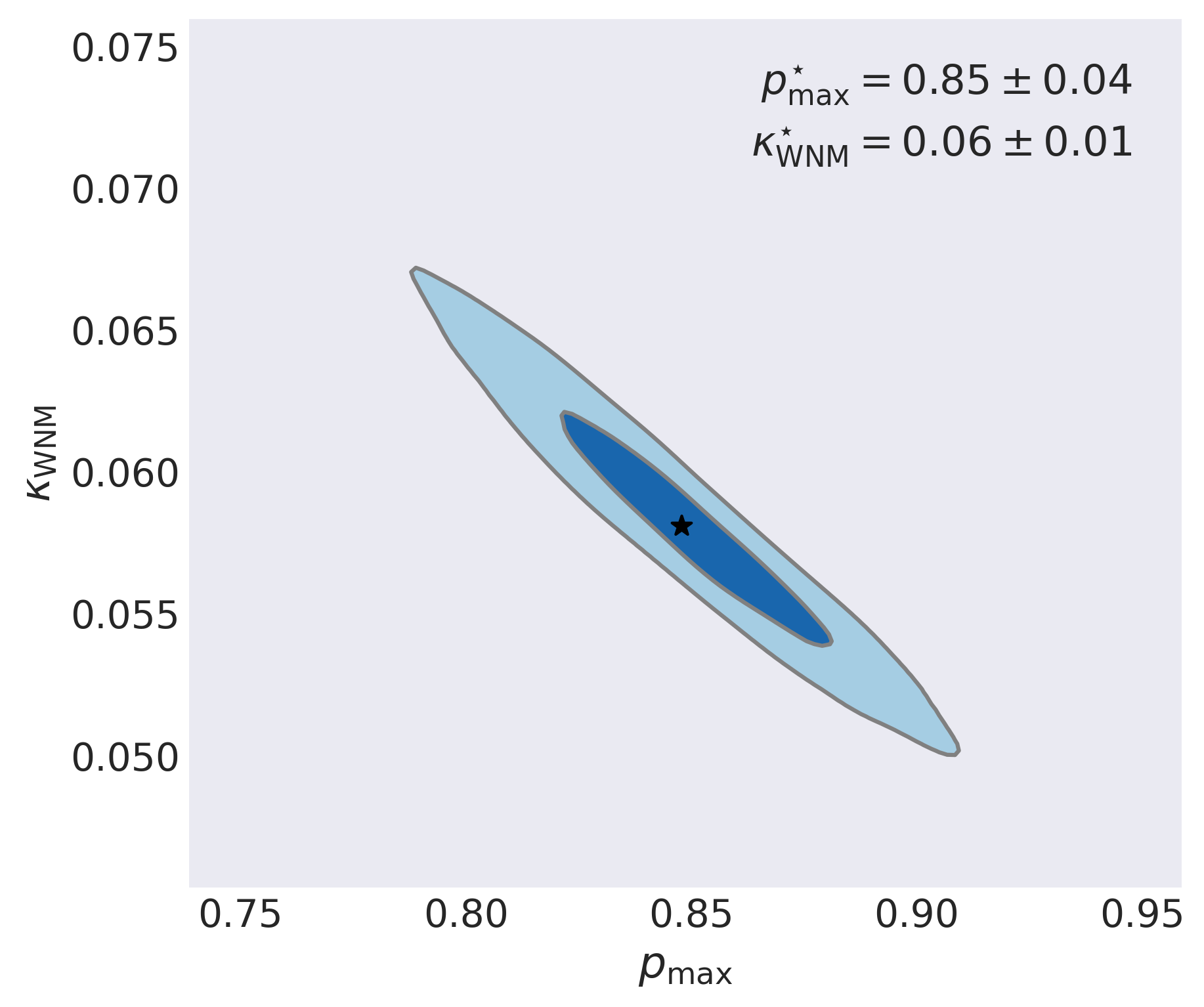}
\caption{Posterior distribution of maximum intrinsic polarization fraction $p_{\mathrm{max}}$ and WNM order parameter $\kappa_{\mathrm{WNM}}$, after fitting the cartoon polarization model described in Figure \ref{fig:cartoon_coherence} to $p_{353}$-$f_{\mathrm{CNM}}$ data. The results show that explaining the observed strong $p_{353}$-$f_{\mathrm{CNM}}$ correlation with a simple model of phase depolarization due to a single disordered WNM component relative to an ordered CNM component requires the intrinsic dust polarization to be unphysically large. }
\label{fig:cartoon_fit_1d}
\end{figure}

\subsection{Phase-Dependent Magnetic Field Interpretation} \label{subsec:phase_coh}

We examine the result of fitting model1 to the observed $p_{353}$ and $f_{\rm CNM}$ data by applying Bayesian hierarchical linear regression. We model the likelihood of the data as 
\begin{align} \label{eq:likelihoods}
    \mathcal{L}&= P(p_{353}, f_{\rm CNM}| \hat{p}_{353}, \hat{f}_{\rm CNM}) P(\hat{p}_{353},\hat{f}_{\rm CNM} | \mathbf{\theta}) \\ \nonumber
    &\propto P(p_{353}| \hat{p}_{353}) P(f_{\rm CNM}| \hat{f}_{\rm CNM}) P(\hat{p}_{353} | \hat{f}_{\rm CNM}, \mathbf{\theta})
\end{align}

where $\hat{p}_{353}$ and $\hat{f}_{\rm CNM}$ are the true, unobserved values of the polarization fraction and CNM fraction, respectively, and $\mathbf{\theta} = (\kappa, p_{\rm max}, \sigma_p^2)$ are the parameters of our model. {We assume standard normal distributions for the likelihood terms}:
\begin{align} \label{eq:prob_pmodel}
P(\hat{p}_{353} | \hat{f}_{\rm CNM}, \mathbf{\theta}) \sim\mathcal{N}({\rm model(\hat{f}_{\rm CNM}, \theta)}, \sigma_p^2)\\
P(p_{353} | \hat{p}_{353}) \sim \mathcal{N}(\hat{p}_{353}, \sigma^2_{\hat{p}_{353}})\\
P(f_{\rm CNM} | \hat{f}_{\rm CNM}) \sim \mathcal{N}(\hat{f}_{\rm CNM}, \sigma^2_{\hat{f}_{\rm CNM}})
\end{align}
where $\rm model(\hat{f}_{CNM}, \theta)$ is given by Equation \ref{eq:plinear}.

We assume a uniform prior distribution for $p_{\rm max}$ and $\kappa_\mathrm{WNM}$ over [0, 1]. For $\sigma_p$, we follow the convention of using a half-Cauchy distribution as the uninformative prior for global variance parameters in Bayesian models \citep{polson11-hc}. Physically, $\sigma_p$ models the additional variance in $p_{353}$ that is uncorrelated with $f_{\rm CNM}$ and thus not captured by model1, e.g. POS dispersion due to inclination angle variation. In Section \ref{sec:corr_interp}, we argued that polarization fraction variation is affected by different factors of grain properties and magnetic field geometry, but the observed $p_{353}$-$f_{\rm CNM}$ relation cannot be explained without an explicit phase-dependent assumption. Here we use model1 to capture the phase-dependent $p$ variation due to relative magnetic field disorder across phases, and encapsulate other sources of $p_{353}$ variance into a global variance parameter $\sigma_p$. 

Performing the Bayesian regression fit with \texttt{PyMC}, we apply Markov Chain Monte Carlo (MCMC) to sample the posterior distribution and show the results for model1 parameters in Figure \ref{fig:cartoon_fit_1d}. The $1\sigma$ best-fit values are $(p_{\rm max}=0.85\pm0.04, \kappa_{\rm WNM}=0.06\pm0.01, \sigma_p=0.036\pm0.001)$. This corresponds to a best-fit slope of 0.80 and intercept of 0.05 for the linear model specified in Equation \ref{eq:plinear}. The relatively small $\sigma_p$ value and $\kappa_{\rm WNM}<1$ are consistent with the discussions in Section \ref{sec:corr_interp} that the observed $p_{353}$-$f_\mathrm{CNM}$ relation cannot be explained without explicit assumption on phase-dependent magnetic field structure variations. $\kappa_{\rm WNM}=0.058$ corresponds to $p_{\rm WNM}=5.8\%$ of $p_{\rm max}$ and therefore highly depolarizing WNM magnetic fields relative to the CNM magnetic fields. However, the best-fit $p_{\rm max}$ value is significantly higher than the observed max $p_{353}\sim0.22$ and far exceeds the maximum intrinsic polarization fractions derived from typical dust grain models \citep{kirchschlager19-ip, draine21-ei}. This means that the current model cannot fully explain the observed $p_{353}$-$f_\mathrm{CNM}$ correlation. Specifically, the model cannot consistently account for both the high slope in the $p_{353}$-$f_{\rm CNM}$ relation and the maximum observed $p_{353}$ value without an unphysically high $p_{\rm max}$. According to Equation \ref{eq:plinear}, the $p$-$f_{\rm CNM}$ slope is $p_{\rm max}(1-\kappa)$. The steep observed $p_{353}$-$f_{\rm CNM}$ slope is such that $p_{\rm max}$ needs to be large, and $\kappa$ and thus $p_{\rm WNM}/p_{\rm CNM}$ must be close to zero. Small $\kappa$ means the WNM column is almost entirely depolarizing. Thus, for the LOS-averaged $p_{\rm model1}$ to match the range of maximum observed $p_{353}\sim0.22$ despite the significant depolarization in the WNM fraction, the small CNM fraction must be highly polarized to make up the difference, leading to the unphysically large $p_{\rm max}=p_{\rm CNM}=0.85\pm0.04$ fit result in model1. On the other hand, if we force $p_{\rm max}=0.22$ with other parameters being the same as the model1 best-fit values, then the resulting $p_{\rm model}$ will only have a 90th percentile value of 0.08, half of the observed 90th-percentile $p_{353}$ value of 0.16. If we instead fit model1 but fix $p_{\rm max}=0.22$, we get best-fit parameters $\kappa_{\rm WNM}=0.39\pm0.01, \sigma_p=0.043\pm0.001)$. The higher $\sigma_p$ value means that more of the $p_{353}$ scatter is attributed factors uncorrelated with $f_{\rm CNM}$, and the resulting $p_{\rm model}$-$f_{\rm CNM}$ relation has a Spearman coefficient of 0.11, much less than 0.50 for the observed $p_{353}$-$f_{\rm CNM}$ correlation. Thus, the observed $p_{353}$-$f_{\rm CNM}$ slope and maximum $p_{353}$ are such that model1 cannot fit the data without an unphysically high $p_{\rm max}$ value. Thus, there must be some additional contribution to the observed $p_{353}$-$f_\mathrm{CNM}$ relationship that is not explained by this simple model of WNM magnetic field tangling relative to an ordered CNM. 

One possible contribution is increased dust emissivity in the CNM. So far, we have implicitly assumed that the dust emissivity is the same between CNM-associated dust and WNM-associated dust. While phase-dependent dust emissivity does not by itself result in a $p_{353}$-$f_\mathrm{CNM}$ relation, if the dust associated with the CNM has a higher emissivity, it will have a higher weight in the LOS average. If the CNM-associated dust also experiences a more ordered magnetic field, this will result in a larger polarization fraction when averaged over the whole column than a simple model where the CNM and WNM components are weighted only by their respective column densities. The result is higher $p$ at the same degree of $f_{\rm CNM}$ compared to when CNM and WNM emissivity are weighted equally, leading to a higher $p_{353}$-$f_\mathrm{CNM}$ slope and alleviating the tension discussed with model1. Past work has found a consistent increase of the FIR emission over \ion{H}{1} column density ratio $I_{857}/N_{\rm H\,I}$ toward sightlines with more CNM \citep{clark19-pn, murray20-cn, lei22-st}. However, since in this study we are restricting to diffuse regions where the variation in the dust to \ion{H}{1} emission ratio is less than 10\% \citep{lenz17-ln}, we do not expect a significant contribution from this effect. For the column density regime ($N_\mathrm{H\,I}<4\times10^{20}~\mathrm{cm^{-2}}$) and scale (80\arcmin) considered in this study, we find that there is only a small $I_{857}/N_{\rm H\,I}$ ratio variation with increasing $f_{\rm CNM}$, consistent with a 10\% scatter. Instead, we reexamine our magnetic field disorder assumptions. 

\begin{figure}[t]
\centering
\includegraphics[width=0.5\textwidth]{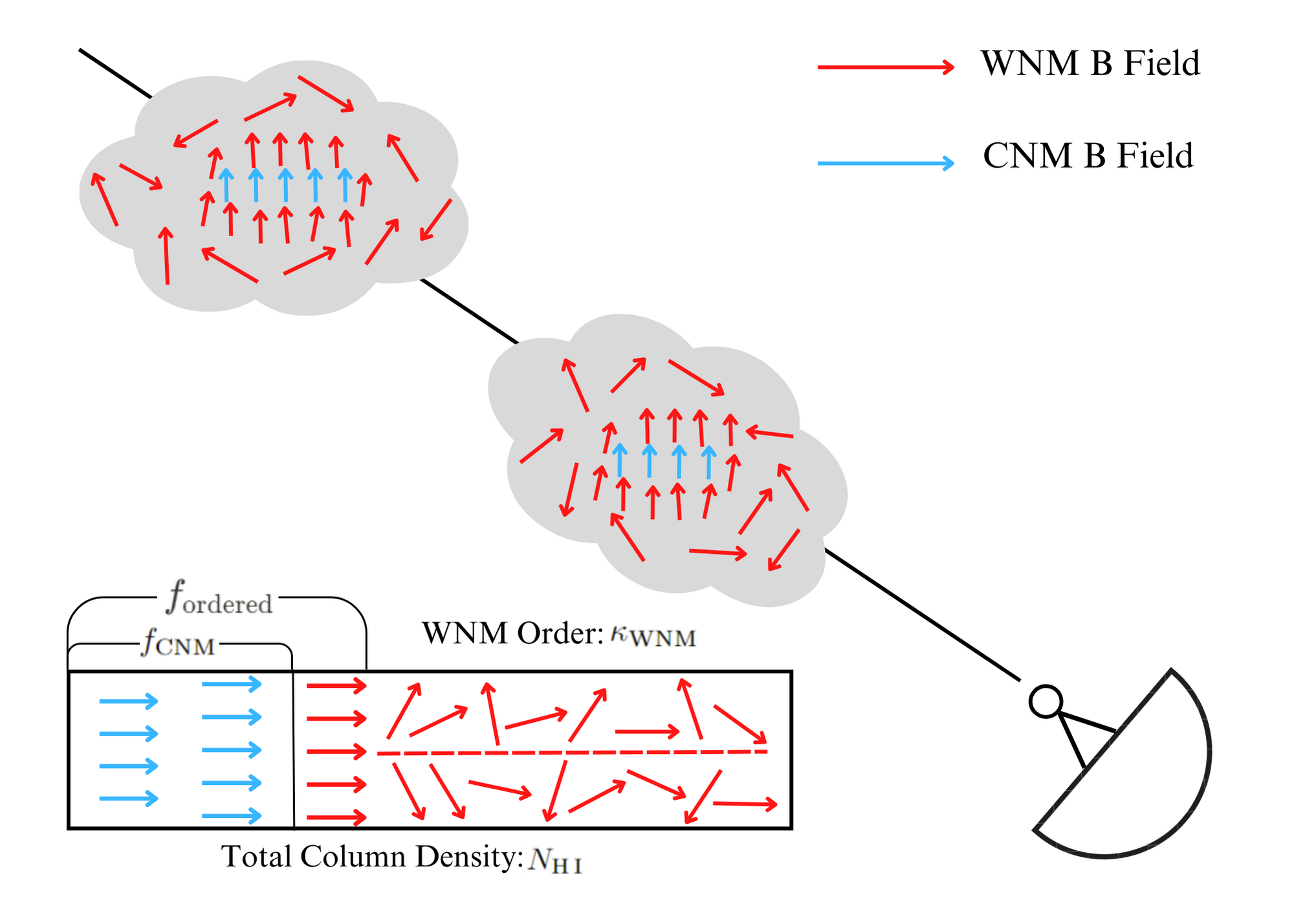}
\caption{Cartoon illustration of the relative disorder of magnetic fields between CNM and WNM as observed along a given line of sight described by model2. Compared to Figure \ref{fig:cartoon_coherence}, we allow for a factor that parameterizes a fraction of the CNM-neighboring WNM region that also has more ordered magnetic field orientations.}
\label{fig:cartoon_coherence2}
\end{figure}

\begin{figure}[t]
\centering
\includegraphics[width=0.45\textwidth]{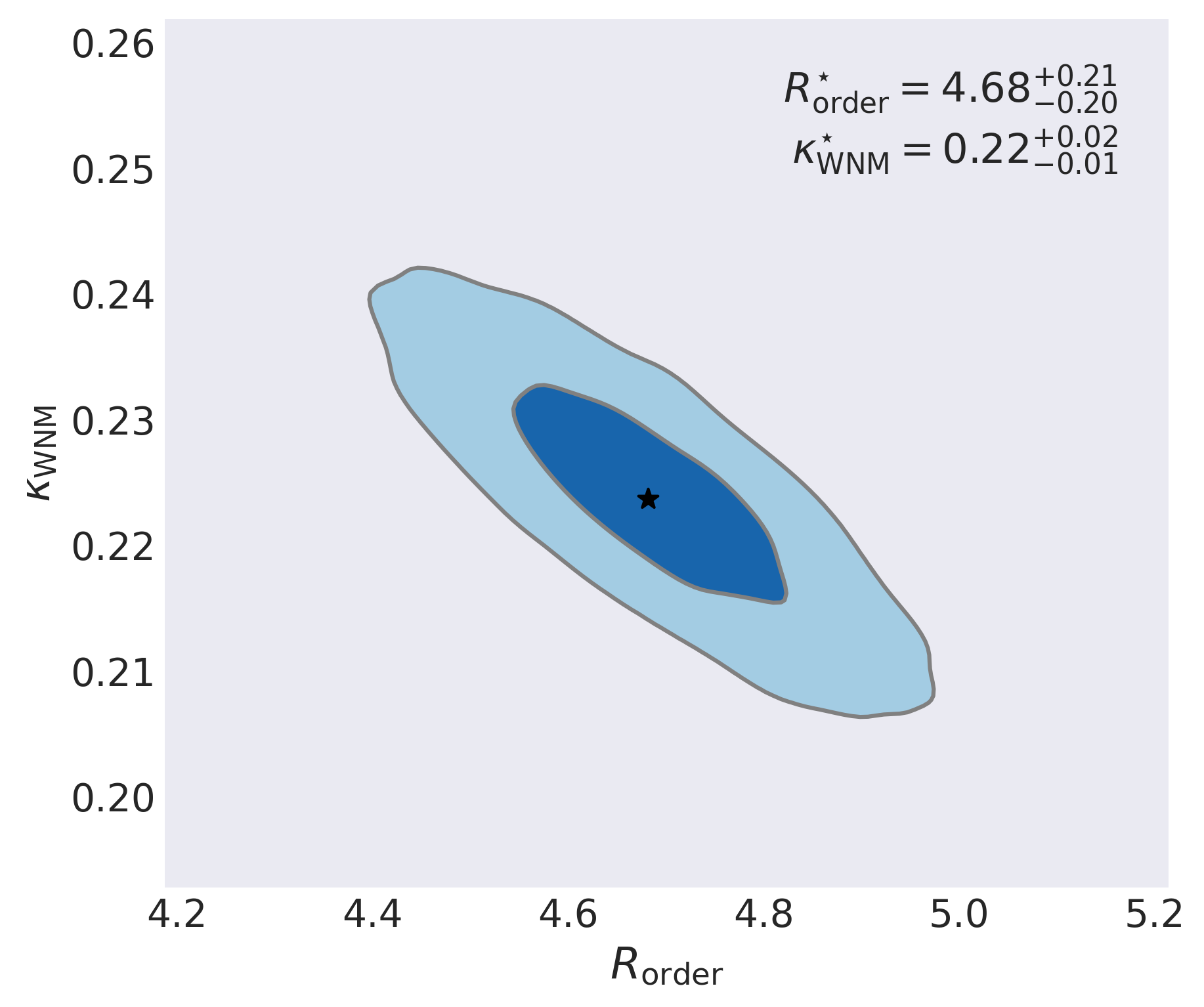}
\caption{Posterior distribution of WNM order parameter $\kappa_{\mathrm{WNM}}$, and ratio of ordered component fraction to CNM fraction $R_{\mathrm{ordered}}=f_{\mathrm{ordered}}/f_{\mathrm{CNM}}$, after fitting model2 to $p_{353}$-$f_{\mathrm{CNM}}$ data. The $1\sigma$ best-fit parameter values are $(R_{\rm order}=4.68^{+0.21}_{-0.20}, \kappa_{\rm WNM}=0.22^{+0.02}_{-0.01})$. }
\label{fig:model2_fit_posterior}
\end{figure}

\begin{figure*}[t]
\centering
\includegraphics[width=0.95\textwidth]{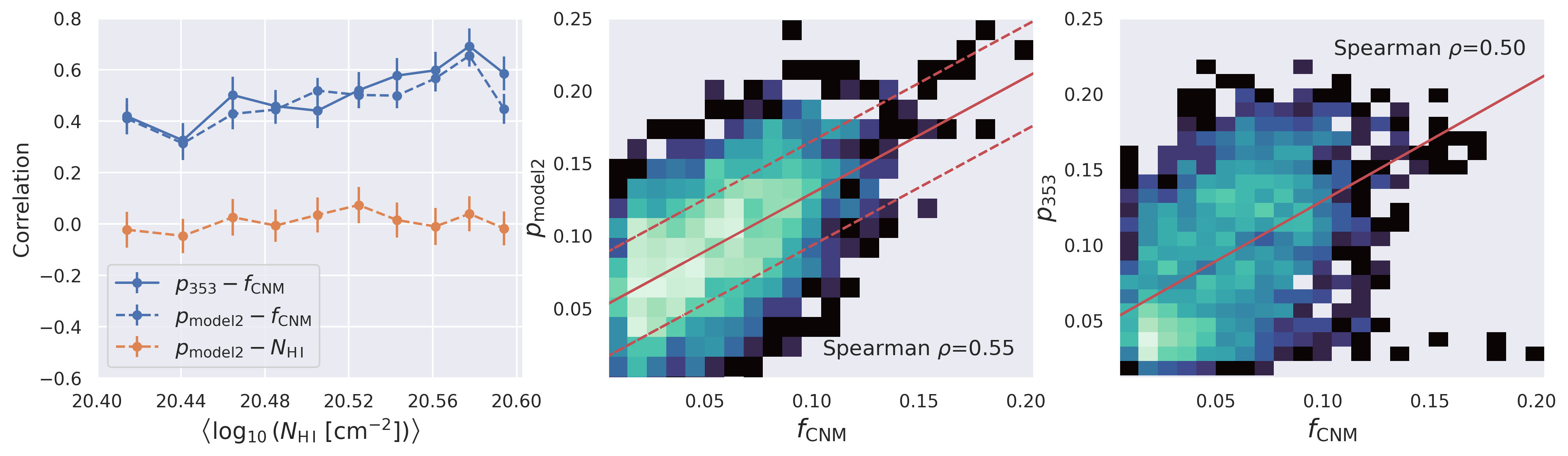}
\caption{Left: Column-density-binned correlation of $f_{\mathrm{CNM}}$ and $N_{\mathrm{H\,I}}$ with model polarization fraction $p_{\mathrm{model2}}$ computed with best-fit parameters ($R_\mathrm{ordered}=4.68$, $\kappa_\mathrm{WNM}=0.22$, $\sigma_p=0.036$). Middle: 2D histogram of $p_{\mathrm{model2}}$-$f_{\rm CNM}$, with the solid red line indicating the best-fit mean $p_{\mathrm{model2}}$ and the dashed red lines the $\sigma_p$ scatter in Equation \ref{eq:prob_pmodel}. Right: The same best-fit linear relation overlaid on the observed $p_{353}$-$f_{\rm CNM}$ 2D histogram. The positive $p_{\rm model2}$-$f_{\rm CNM}$ correlation is consistent with the observed $p_{353}$-$f_{\rm CNM}$ correlation, while $p_{\rm model2}$-$N_{\rm HI}$ is consistent with no correlation. }
\label{fig:pmodel2_fcnm_corr}
\end{figure*}

In the cartoon model illustrated in Figure \ref{fig:cartoon_coherence}, we assumed that along each sightline there is a single parameter $\kappa_\mathrm{WNM}$ characterizing the tangling of the WNM (i.e., the non-CNM) magnetic field relative to that of the CNM. In general, in the canonical picture of a thermally bistable \ion{H}{1} gas, the CNM condenses out of the diffuse WNM driven by turbulence and thermal instability \citep{wolfire03-na, saury14-na}. If the WNM magnetic field is more disordered to begin with, and a more ordered CNM magnetic field forms out of the compression and condensation process, then we might naturally expect the gas that neighbors the sites of CNM formation and the thermally unstable medium to have intermediate degrees of magnetic field disorder as well. We could further hypothesize that the fraction of this ``ordered WNM" region associated with CNM formation would correlate with $f_\mathrm{CNM}$. This would enhance the effect of a more ordered CNM magnetic field adding constructively to higher observed $p$, resulting in a stronger $p_{353}$-$f_\mathrm{CNM}$ correlation. This scenario is illustrated in the updated cartoon model shown in Figure~\ref{fig:cartoon_coherence2}. We denote the updated model ``model2" in contrast to model1 presented in Figure \ref{fig:cartoon_coherence}.

Here, in the spirit of the cartoon model, instead of trying to model the exact distribution of this variation, we define $f_\mathrm{ordered}$ to parameterize the total fraction of magnetically ordered regions including both the CNM and the fraction of the WNM column we assume to be ordered. To encode that the additional ordered component should correlate with $f_\mathrm{CNM}$, we define $R_\mathrm{ordered}$ as a ratio of $f_\mathrm{ordered}$ and $f_\mathrm{CNM}$:
\begin{equation}
    R_\mathrm{ordered}=f_\mathrm{ordered}/f_\mathrm{CNM}.
\end{equation}
We emphasize that the fact that $f_\mathrm{ordered}$ is proportional to $f_\mathrm{CNM}$ means that this model is not degenerate with simply raising $\kappa$ in model1. Instead, $R_\mathrm{ordered}$ serves as an additional weighting factor to the slope of the $p_{533}-f_{\rm CNM}$ relation. We introduce $R_\mathrm{ordered}$ as a new parameter in model2. Furthermore, instead of fitting $p_{\rm max}$ as a free variable, we fix it to a physically motivated value. As a result, model2 is a re-parameterization of the linear model of the $p_{533}-f_{\rm CNM}$ relation from ($p_{\rm max}$, $\kappa_{\rm WNM}$) in model1 to ($R_\mathrm{ordered}$, $\kappa_{\rm WNM}$), and Equation \ref{eq:plinear} becomes:
\begin{align} \label{eq:plinear2}
    p_{\rm model2} &= f_\mathrm{ordered}\cdot p_{\rm max} + (1- f_\mathrm{ordered})\cdot p_{\rm WNM} \\ \nonumber
                   &= p_{\rm max}\bigl(Rf_{\rm CNM}+(1- Rf_{\rm CNM})\kappa\bigr) \\ \nonumber
                   &= p_{\rm max}R(1-\kappa)f_{\rm CNM}+p_{\rm max}\kappa
\end{align}
We fix $p_{\rm max}$ according to the maximum observed $p^{\rm max}_{353}=22^{+3.5}_{-1.4}\%$ as determined by Planck full-sky observations at 353 GHz and $80\arcmin$ resolution \citep{planck18-cr}. The state-of-the-art model of interstellar dust grains known as Astrodust \citep{hensley22-ad} gives a similar constraint at $p_{\rm max}\sim19.2\%$ at 353 GHz. The uncertainty on $p^{\rm max}_{353}$ is dominated by uncertainty on the total intensity zero level. Here we present the result for setting $p^{\rm max}_{353}=22\%$, assuming the fiducial Galactic emission offset of 40 $\mu\mathrm{K_{CMB}}$ \cite{planck18-cr}. We also repeat the same analysis using the upper and lower limits for this offset. 

Performing the Bayesian regression fit process described by Equations \ref{eq:likelihoods}-\ref{eq:prob_pmodel} for model2, replacing the Equation \ref{eq:plinear} parameterization of the linear relation with Equation \ref{eq:plinear2}, we show the resulting posterior distributions of model2 parameters in Figure \ref{fig:model2_fit_posterior}. Since we are still performing the same linear model fit just with the slope and intercept reparametrized, the (slope, intercept, $\sigma_p$) has the same best-fit values of (0.80, 0.05, 0.036) as in model1. The $1\sigma$ best-fit values of the model2 parameters are $(R_{\rm order}=4.68^{+0.21}_{-0.20}, \kappa_{\rm WNM}=0.22^{+0.02}_{-0.01})$. In the diffuse regions we are considering, the CNM accounts for $\sim5\%$ of the total \ion{H}{1} mass \citep{murray20-cn}. Thus, $R_\mathrm{ordered}=4.68$ would imply that the magnetically ordered fraction of the WNM column accounts for an additional $\sim18.4\%$ of the \ion{H}{1} mass. Using the best-fit ($R_\mathrm{ordered}=4.68$, $\kappa_\mathrm{WNM}=0.22$, $\sigma_p=0.036$) parameters with $p_{\rm max}=0.22$, we make a random realization of $p_{model2}$ using Equations \ref{eq:prob_pmodel} and \ref{eq:plinear2} and show its correlation relations with observed data in Figure \ref{fig:pmodel2_fcnm_corr}. $R_\mathrm{ordered}$ and $\kappa_\mathrm{WNM}$ determine the best-fit mean $p_{model2}$ line shown in red, while the scatter in $p_{model2}$ is due entirely to $\sigma_p$. We verify that the degree of $p_{\rm model2}$-$f_{\rm CNM}$ correlation is consistent with the observed $p_{353}$-$f_{\rm CNM}$ behavior, while $p_{\rm model2}$-$N_{\rm HI}$ is compatible with no correlation. These results show that the cartoon model of magnetic field structure variation across phases illustrated in Figure \ref{fig:cartoon_coherence2} is compatible with the observed $p_{353}-f_\mathrm{CNM}$ correlation. 

We repeat the modeling fitting procedure while adopting the high vs. low Galactic offset limit to explore the effect of total intensity zero-level uncertainty on the fitting results. Following \citet{planck18-cr}, we apply a low offset of 23 $\mu\mathrm{K_{CMB}}$ with $p^{\rm max}_{353}=25.5\%$, and a high offset of 103 $\mu\mathrm{K_{CMB}}$ with $p^{\rm max}_{353}=20.6\%$. The best-fit values are $(R_{\rm order}=4.13^{+0.20}_{-0.21}, \kappa_{\rm WNM}=0.22^{+0.02}_{-0.01})$ and $(R_{\rm order}=4.74^{+0.19}_{-0.20}, \kappa_{\rm WNM}=0.21^{+0.02}_{-0.01})$ for the low and high offset respectively, consistent with the fit with fiducial offset values to within $\sim12\%$. Note that best-fit $R_{\rm order}$ increases more significantly with decreased fixed $p_{\rm max}$ value while $\kappa_{\rm WNM}$ barely changes. Thus in this model, a lower polarization fraction in the CNM, which may be more easily explained by physical dust models, requires a larger $f_{\rm CNM}$-correlated dust column to be polarized at the same level. {Finally, we also explored the potential spatial variation of the parameters by refitting the model to subsets of our data. Fitting the model to two halves of the dataset randomly drawn from our region of interest, we find that the best-fit $\kappa$ and $R_{\rm ordered}$ agree between datasets within uncertainties. We also repeated the modeling fitting to the fraction of \ion{H}{1}4PI dataset that has a similar $f_{rm CNM}$ dynamic range as the GALFA-\ion{H}{1} region of interest analyzed in this section, and find the best-fit model parameters to be $(R_{\rm order}=3.20^{+0.20}_{-0.22}, \kappa_{\rm WNM}=0.23^{+0.01}_{-0.02})$, qualitatively consistent with the phase-dependent magnetic field geometry picture presented here. Further discussion of spatial variation can be found in Appendix \ref{appx:galfa_v_hi4pi}.}

\begin{figure*}[t]
\centering
\includegraphics[width=\textwidth]{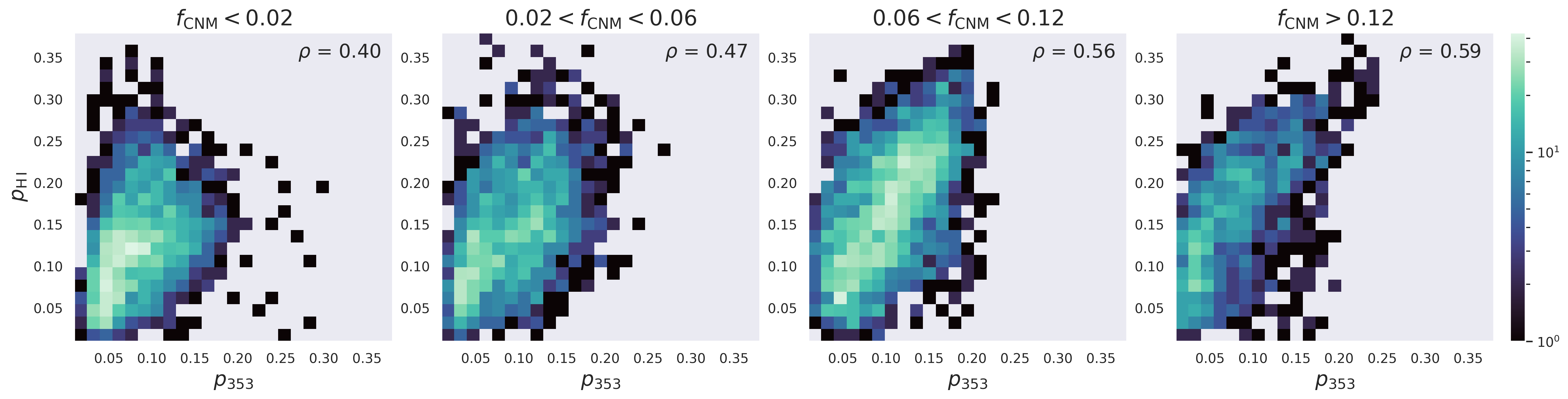}
\caption{Two-dimensional histograms of $p_{353}$ and $p_{\mathrm{H\,I}}$ in bins of $f_{\mathrm{CNM}}$. The $p_{353}$-$p_{\mathrm{H\,I}}$ correlation increases with increasing $f_{\mathrm{CNM}}$, suggesting that the \ion{H}{1} polarization template traces the dust polarization fraction better at higher CNM fraction. }
\label{fig:p353_phi_corr}
\end{figure*}

\begin{figure}[t]
\centering
\includegraphics[width=0.45\textwidth]{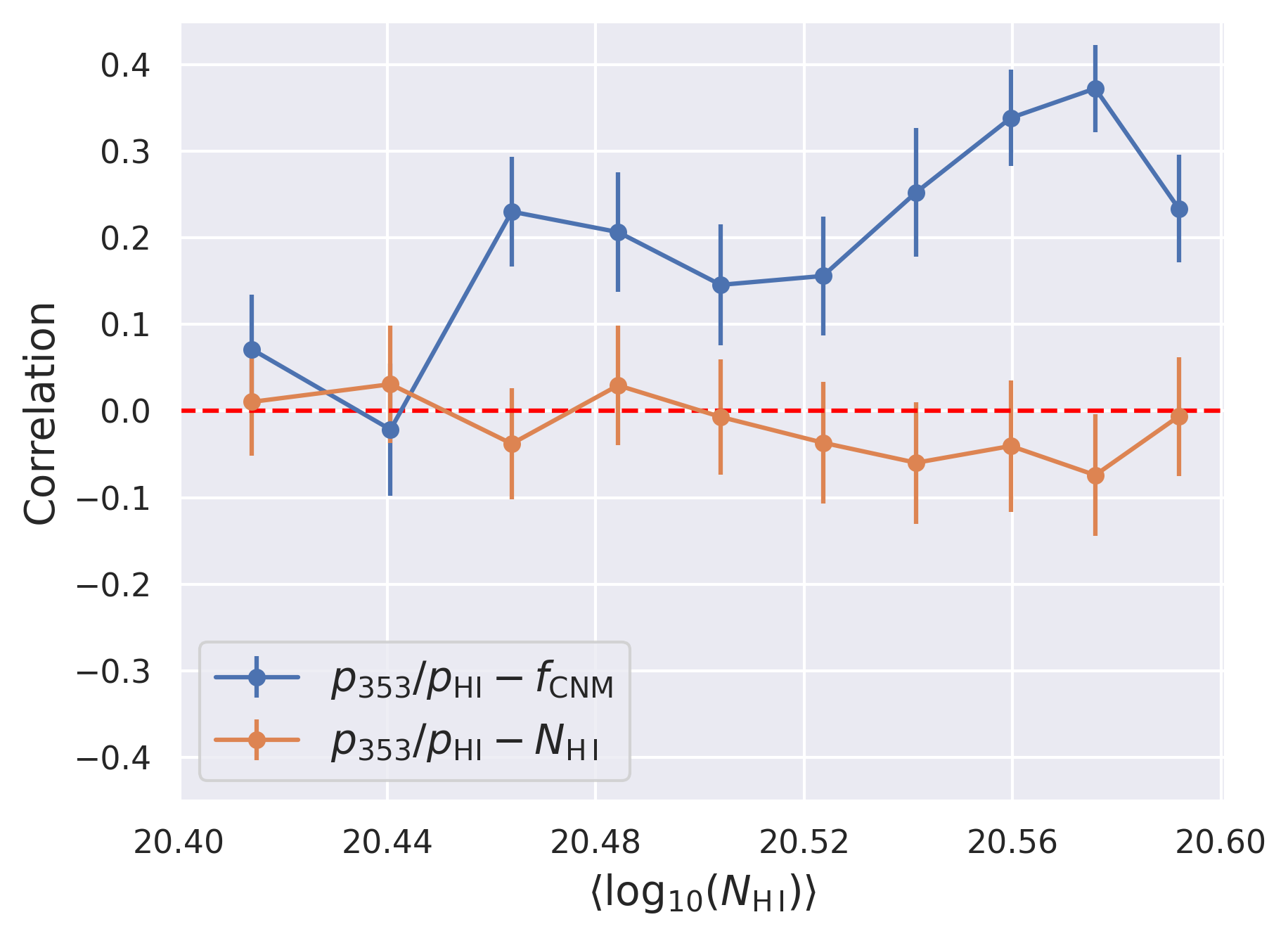}
\caption{Column-density-binned correlation of polarization fraction ratio $p_{353}/p_{\mathrm{H\,I}}$ with $f_{\mathrm{CNM}}$ and $N_{\mathrm{H\,I}}$. There is a significant positive $p_{353}/p_{\mathrm{H\,I}}$-$f_{\mathrm{CNM}}$ correlation while correlation with $N_{\mathrm{H\,I}}$ is compatible with 0. This is consistent with the interpretation that the $p_{353}$-$p_{\mathrm{H\,I}}$ difference can be partly attributed to the \ion{H}{1} polarization template overweighting a more ordered CNM component and underweighting WNM tangling. The agreement improves with increasing CNM fraction and $p_{353}$ increases relative to $p_{\mathrm{H\,I}}$ as WNM tangling plays a lesser role. }
\label{fig:p353_phi_ratio}
\end{figure}

\subsection{Consistency with \ion{H}{1} Polarization Template} \label{subsec:hi_pol_int}
As described in Section \ref{subsec:hi_pol}, using the orientation of filamentary structures that trace local magnetic fields, \citet{clark19-hs} constructed 3D (position-position-velocity) Stokes polarization parameter maps from \ion{H}{1} intensity data. The \ion{H}{1}-based $Q$ and $U$ maps integrated over LOS velocity are found to be well-correlated with the 353 GHz dust $Q$ and $U$ maps. However, since the \ion{H}{1}-based polarization templates are constructed from preferentially CNM structures \citep{kalberla18-cw, clark19-pn, peek19-sc, murray20-cn}, we expect them to overweight the contribution of the CNM to the total polarized emission. We examine the variation of the correlation between the \ion{H}{1}-based and 353 GHz dust map in regions binned by $f_\mathrm{CNM}$ values, to test the consistency of the phase-dependent magnetic field variation interpretation discussed in the previous section. 

\citet{clark19-hs} find that the \ion{H}{1}-based polarization fraction $p_\mathrm{H\,I}$ is well-correlated with $p_{353}$ \citep[see also][]{clark18-np}. However, if the \ion{H}{1} polarization template over-weights CNM structures, and if there is a difference in CNM and WNM magnetic field disorder, we should expect the $p_\mathrm{H\,I}$-$p_{353}$ correlation to be stronger in regions with higher CNM content. In Figure \ref{fig:p353_phi_corr}, we plot the distribution of $p_\mathrm{H\,I}$ vs. $p_{353}$ in regions of increasing $f_\mathrm{CNM}$, and find a modest but consistent trend of stronger correlation with increasing CNM fraction. The Spearman correlation coefficient improved from $\rho=0.40$ at $f_\mathrm{CNM}<0.02$ to $\rho=0.59$ at $f_\mathrm{CNM}>0.12$. The qualitative behavior of this correlation trend is insensitive to the choices of $f_\mathrm{CNM}$ range. Furthermore, if the WNM magnetic field orientation is more disordered relative to that of the CNM as proposed in the previous section, by over-weighting a more ordered CNM, $p_\mathrm{H\,I}$ should in general overestimate $p_{353}$ in low $f_\mathrm{CNM}$ regions relative to high $f_\mathrm{CNM}$ regions. This should translate to a positive correlation between the $p_{353}$/$p_\mathrm{H\,I}$ ratio and $f_\mathrm{CNM}$. In Figure \ref{fig:p353_phi_ratio}, we compute the $p_{353}$/$p_\mathrm{H\,I}$-$f_\mathrm{CNM}$ correlation in the same equal-area column density bins used for $p_{353}$-$f_\mathrm{CNM}$ correlation in Figure \ref{fig:p353_fcnm_corr_select}. We observe a consistent positive correlation as expected for the physical picture we put forward in model2 over most of the column density range, except in the two lowest column density bins. Therefore, the results in Figures \ref{fig:p353_phi_corr} and \ref{fig:p353_phi_ratio} together are consistent with a phase-dependent magnetic field variation interpretation where the magnetic field in the WNM is disordered relative to that of the CNM, resulting in a positive correlation of polarization fraction with CNM content. 

\section{Discussion} \label{sec:discussion}

\subsection{Interpretations of Phase-Dependent Magnetic Field Properties} \label{subsec:phase_coh_imp}
The results presented in Sections \ref{sec:corr_interp} and \ref{sec:corr_model} show that a difference in magnetic field tangling between the WNM and the CNM is the most likely explanation for the observed positive $p_\mathrm{353}$-$f_\mathrm{CNM}$ correlation. Thus the results presented here constitute a new constraint on the relationship between multi-phase ISM structure and the magnetic field. Our work complements investigations of magnetic field alignment between ionized and neutral phases, which have mostly focused on small patches of the sky where a direct morphological connection between different tracers is found \citep{zaroubi15-lh, jelic18-lh, bracco20-lh, campbell22-cm}. Here we present a {new} study constraining the relative disorder of the magnetic field between neutral phases of the ISM over large regions of the sky, using statistics of the dust polarization fraction and phase-decomposed maps of \ion{H}{1} emission. With the development of ever-more sophisticated phase decomposition techniques \citep{marchal19-rs, riener20-ag, murray20-cn}, alongside the increasing availability of \ion{H}{1} absorption sightlines from future and ongoing surveys \citep{dickey13-gs, McClure-Griffiths:2015ZO}, there are growing opportunities to further test this picture to examine the important question of magnetic field alignment between phases in different environments across different scales. 

The data-driven cartoon dust emission model presented in \ref{subsec:phase_coh} allows us to quantitatively constrain the properties of the CNM and the WNM magnetic field. The model parameter posterior distributions in Figure \ref{fig:model2_fit_posterior} suggest that a CNM with more aligned magnetic fields forms out of the WNM with generally disordered fields. In particular, the best-fit value $\kappa_\mathrm{WNM}\sim0.22$ describes the degree of fractional depolarization due to magnetic field tangling in the WNM column relative to the CNM column. How should we interpret this different degree of geometrical depolarization in terms of the 3D magnetic field structure in each phase? The difference in magnetic field disorder in the WNM and CNM could arise from two main physical pictures. On the one hand, a single statistical distribution of magnetic field structure could exist independent of \ion{H}{1} phase, but the magnetic field in the CNM column could be more ordered because the CNM is confined to a much smaller path length and thus samples the distribution only at that smaller scale. The typical CNM path length is $\sim$pc scale, while the WNM is more volume filling with a typical path length of 100 pc or more \citep{heiles03-ml, kalberla09-hi}. The $f_{CNM} - p_{353}$ correlation could then be driven by the higher density of the CNM gas. In this case, our measurement of the difference in CNM and WNM magnetic field disorder constrains the overall LOS magnetic field tangling over the WNM scale vs. the CNM scale, and could be used to constrain the scale dependence of the 3D magnetic field structure in the neutral medium. 

On the other hand, the difference in magnetic field disorder between the CNM and the WNM could point to distinct magnetic field distributions between the phases. {Specifically, the CNM magnetic field is more ordered over CNM path lengths than the WNM magnetic field over the WNM path length.} We argue that the observational constraints presented in this study favor this interpretation {over a picture in which the magnetic field geometry is completely independent of the gas phase structure, such that the CNM column simply samples} the same magnetic field distribution over different scales than the WNM column. First, the correlation between dust polarization angle and \ion{H}{1} CNM filaments implies that in the diffuse ISM, the mean magnetic field orientation in the CNM is well-aligned with the mean magnetic field orientation in the WNM. The correlation between $p_{353}$ and $f_{\rm CNM}$ then constrains the relative dispersion of the WNM magnetic field from the mean orientation. As the middle panels of Figure \ref{fig:p353_fcnm_lim} show, a strong degree of $p_{353}$-$f_{\rm CNM}$ correlation persists in regions with almost perfect $\theta_{H\,I}$-$\theta_{353}$ alignment ($\xi>0.95)$ on the 80\arcmin\ scales considered in this work. 
If we argued that the scale-dependence alone were responsible for the different magnetic field dispersion attributed to the CNM and the WNM, we would need to consider the plausibility of a magnetic field geometry that has the WNM-measured dispersion over $\sim100$ pc scales, but which on any given $\sim$pc-scale region would appear both ordered \textit{and} aligned with the WNM mean magnetic field orientation. Future work should further explore the different interpretations through detailed study of the magnetic field distribution in CNM formation simulations \citep{inoue16-fh, kim17-ti, gazol21-pg, Moseley21-td, fielding23-pm}. 

A further constraint is the large best-fit value of $R_{\rm ordered}\sim4.68$, indicating that a significant portion of the non-CNM column also has a relatively ordered magnetic field. If the additional ordered column spans a significant path length, then the tension between the mean-field constraint and the relative dispersion constraint described above becomes even more severe. Thus the combination of the CNM and this additional ordered component of the column is likely to have a statistically more ordered 3D magnetic field distribution than the disordered WNM column. For the diffuse region considered in this study, $R_{\rm ordered}\sim4.68$ translates to an average \ion{H}{1} mass fraction of $\sim18.4\%$ for the additional ordered non-CNM column. A natural physical origin for this additional column could be the UNM. Utilizing absorption measurements from the 21-SPONGE survey, \citep{murray18-sp} found the UNM mass fraction to be generally consistent with $\sim20\%$ of total \ion{H}{1} mass \citep{murray18-sp}, with the caveat that the region surveyed was not necessarily representative of the diffuse sky we are considering. Ongoing and future \ion{H}{1} absorption surveys will add to available data at high latitudes and better constrain properties of the UNM in these regions \citep{dickey13-gs, McClure-Griffiths:2015ZO, dickey22-ga}. If the additional magnetically ordered column is UNM, assuming the UNM has a density between the fiducial WNM and CNM densities $[\rho_{\rm CNM}=0.6{\rm\,cm^{-3}},\rho_{\rm WNM}=30{\rm\,cm^{-3}}]$ \citep{draine11-bk}, then the ordered column on average spans a path length between [1pc, 40pc], further adding to the tension between the mean-field constraint and the relative dispersion constraint that any model without an explicitly phase-dependent magnetic field structure needs to explain. 

An alternative to interpreting $R_{\rm ordered}$ as the additional magnetically ordered column is the possible role of dust emissivity variation. Higher emissivity in the magnetically ordered regions would lead to higher weighting for the ordered components in the LOS average. Since we modeled $R_\mathrm{ordered}$ as a ratio $f_{\rm ordered}/f_{\rm CNM}$, $R_\mathrm{ordered}$ effectively serves as an additional weighting term to the CNM contribution in Equations \ref{eq:qm1} and \ref{eq:um1}, degenerate with the effect of a higher dust emissivity weighting. Hence, assuming CNM emissivity $\epsilon_{\rm CNM}$ $\geq$ WNM emissivity $\epsilon_{\rm WNM}$, we should consider $f_\mathrm{ordered}\sim4.68 f_{\rm CNM}$ as an upper limit on the fraction of magnetically-ordered WNM column. However, we do not consider emissivity variation a significant effect in this study because of the small variation ($\sim$10\%) of the dust emission to \ion{H}{1} column ratio in the diffuse sky \citep{lenz17-ln}. A significant source of the scatter in that relationship in these regions is the photometric uncertainties. However, even if the scatter is caused entirely by higher emissivity in the CNM, in the range of $f_{\rm CNM}\sim(0, 0.2)$ considered in our study, a ~10\% scatter in the dust emission to \ion{H}{1} column ratio would only correspond to an emissivity ratio $R=\epsilon_{\rm CNM}/\epsilon_{\rm WNM}\sim1.5$ relative to the WNM, outside the $2\sigma$ best-fit range we found for $R_{\rm ordered}$.

In summary, the observational constraints presented in this study are most consistent with a higher degree of magnetic field orientation dispersion in the WNM than in the CNM. Numerical simulation studies of CNM formation have mostly focused on the physical and morphological properties of the CNM magnetic field, such as the alignment of magnetic field orientation with cold filaments \citep{inoue16-fh, gazol21-pg}. Our results on the relative disorder of the magnetic field between the CNM and the WNM columns place a new constraint on CNM formation models that can be directly compared against MHD simulations.

\subsection{Comparison with Other Work} \label{subsec:compare_work}
{Other work has investigated the magnetic field structures between different neutral ISM phases. In particular, \citet{ghosh17-ms} (hereafter G17) and \citet{adak20-dp} (hereafter A20) proposed a dust model that incorporates the CNM, UNM, and WNM-associated emission as three discrete layers. Fitting to one- and two-point dust polarization statistics derived from Planck observations, the authors conclude that the magnetic field in the CNM is more turbulent than in the UNM/WNM component.} 

{Our work differs from the previous papers in two major ways. First, G17 and A20 fit the strength of the turbulent magnetic field relative to an ordered mean field in the POS over the scale of the Northern and Southern Galactic cap regions. Along the LOS, the three phases CNM, UNM, and WNM are modeled as three discrete layers with no internal LOS magnetic field variations. In this work, we study the relative depolarization in the CNM vs. WNM column, which corresponds physically to depolarization over the typical CNM and WNM path lengths along the LOS. Fitting to the $p_{353}$-$f_{\rm CNM}$ correlation does not directly constrain the coherence of the POS magnetic field over a scale larger than the 80\arcmin\ beam size of our data. Thus our inference of a more ordered CNM magnetic field along the LOS does not necessarily conflict with the POS dispersion in the CNM layer found in the G17 and A20 works.} 

{Furthermore, while we utilize a phase-decomposed CNM mass fraction map directly, G17 and A20 treat phase decomposition as part of their model fit. The resulting column density maps for the three phases in Figure 1 of G17 show that their inferred CNM column density is often significantly higher than the combined UNM and WNM column densities even in very high Galactic latitude areas, indicating a CNM fraction $>0.4$ across most regions. However, \citet{murray18-sp} absorption measurements show that the CNM fraction over the high latitude ($|b|>30\degree$) region is significantly lower, with only 5 out of the 58 sightlines in that work having $f_{\rm CNM}>0.4$. The data-derived CNN-based $f_{\rm CNM}$ map used in this work agrees well with the measured data at the absorption sightlines, and shows a mean $f_{\rm CNM}\sim0.1$ over the whole GALFA-\ion{H}{1} high latitude region and minimal CNM content at $|b|>60\degree$ \citep{murray20-cn}. The discrepancy between the column attributed to the CNM component makes it difficult to compare the conclusions in these works quantitatively. Future studies incorporating phase-decomposed \ion{H}{1} maps into a dust model fit to one- and two-point polarized dust emission statistics could enable a more direct comparison with these works.} 


\subsection{Column-Density-Dependent Correlation Behavior} \label{subsec:col_den_var}
In this study, we focus on exploring interpretations for the positive $p_\mathrm{353}$-$f_\mathrm{CNM}$ correlation in the most diffuse regions of the sky. As Figures \ref{fig:p353_fcnm_corr} and \ref{fig:compare_col_den} show, the full $p_\mathrm{353}$-$f_\mathrm{CNM}$ relation transitions from positive correlation to anticorrelation at higher column densities ($N_\mathrm{H\,I}\sim10^{21}\ \mathrm{cm^{-2}}$). The explanation for this transition, and whether the phase-dependent magnetic field variation interpretation is consistent with the anticorrelation at high $N_\mathrm{H\,I}$, are important questions that should be explored in future work. Factors ruled out in the most diffuse regions might play a more important role in dust depolarization at higher column densities, such as the loss of grain alignment efficiency and multiple LOS \ion{H}{1} components. For LOS complexity, at higher column densities and higher $f_\mathrm{CNM}$ values, it is more likely that the CNM structures are distributed across separate \ion{H}{1} clouds, and the resulting dispersion across multiple components would lead to strong depolarization from within the CNM-associated dust column. To test that hypothesis in future studies, we would need to measure the number of CNM components per sightline and their associated column densities. {On the simulation side, past work has explored magnetic field dispersion of cold gases mainly in the high column density regimes where we observe $p_\mathrm{353}$-$f_\mathrm{CNM}$ anticorrelation. Modeling the formation of cold dense clouds in the column density regime of $N_\mathrm{H\,I}\sim10^{21}\ \mathrm{cm^{-2}}$, \citet{kritsuk17-si} found a higher Alfv\'{e}nic Mach number in the cold gas compared to the warm gas. This would imply a higher depolarization in the CNM than the WNM due to dispersion and naturally leads to a $p$-$f_\mathrm{CNM}$ anticorrelation consistent with what we observe in this column density regime. The implication of our work that we should observe a qualitatively different behavior at lower column density regimes should be explored further in future studies using CNM formation simulations \citep{inoue16-fh, kim17-ti, gazol21-pg, Moseley21-td, fielding23-pm}.}

\subsection{Implications for Dust Grain Models}
The level of maximum observed polarization fraction $p^{\rm max}_{353}\sim20\%$ found by Planck observations \citep{planck14-pd, planck18-cr} are challenging to reproduce for dust models \citep{draine09-dm, guillet18-dm, hensley21-dm}. Characterizing optical starlight polarization in regions of maximally polarized dust emission, \citet{panopoulou19-dm} find that the high polarization fractions are unlikely to result from dust properties such as enhanced grain alignment. Instead, they argue that a favorable magnetic field geometry is the most likely explanation, where in the regions of maximum polarization, the magnetic field is mostly in the plane of sky and uniform along the line-of-sight. Here we argue that our results regarding magnetic field disorder between neutral phases, added to the favorable magnetic field geometry argument, potentially make it easier to reconcile the high maximum polarization observation with dust models.  

Any sightline will suffer from some degree of geometric depolarization both along the line of sight and within the beam. This is certainly the case at the resolution of the Planck dust maps of 80\arcmin\ beam size over typical WNM path lengths of over $\sim 100$pc. But if the CNM column is more magnetically ordered than the WNM column, and high polarization fraction regions are associated more with high CNM content which contributes much more significantly to the LOS-averaged $p$ value than the WNM column, then the relevant scale for the favorable magnetic field geometry argument will be the much smaller CNM path length scale $\sim1$pc. It is more reasonable to expect minimal geometric depolarization over the small and more magnetically ordered CNM columns, strengthening the explanation that favorable magnetic field geometry brings the observed maximum polarization fraction closer to the theoretical maximum. 

\subsection{Implications for Dust Foreground Modeling for Cosmology} \label{subsec:hi_imp}
In section \ref{subsec:hi_pol_int}, we explored the correlation of the \ion{H}{1}-based polarization template by \citet{clark19-hs} with Planck 353 GHz maps, and found that the variation is consistent with magnetic field disorder between the CNM and the WNM. The variation of the $p_{353}$-$p_{\rm HI}$ correlation and the $p_{353}/p_{\rm HI}$ ratio in regions of differing CNM fraction suggests a path to improve the template in the future, e.g. by taking into account CNM fraction weighting using phase-decomposed \ion{H}{1} maps. 

In general, the complexity of the dust and magnetic field distributions along the LOS is a major source of uncertainty in dust foreground modeling for cosmic microwave background (CMB) polarization experiments \citep{tassis15-ls, martinez18-ls, mcbride23-ls, vacher23-ls}. Compared to \citet{clark19-hs}, other \ion{H}{1}-based CMB foreground dust polarization models have approached the problem by assuming discrete layers of LOS dust components, either as a free parameter \citep{planck16-dl}, or to represent phase composition \citep{ghosh17-ms, adak20-dp}. For the discussion of phase composition, \citet{ghosh17-ms} and \citet{adak20-dp} model each phase as one discrete layer without any internal magnetic field structure variation along the LOS, and fit their model to one- and two-point statistics of the polarized dust emission without explicitly considering its relation with phase content. The variation of polarized dust emission with LOS complexity $N_c$ \citep{panopoulou20-nc} and CNM mass fraction $f_{\rm CNM}$ shown here demonstrates that simultaneously fitting the dust polarization and \ion{H}{1} data requires physically motivated modeling of magnetic field structure variations across LOS, multi-phase \ion{H}{1} components. 

\section{Summary and Conclusions} \label{sec:conclusions}
In this study, we examined the correlation of Planck dust polarization fraction at 353 GHz $p_{353}$, CNM mass fraction $f_\mathrm{CNM}$ \citep{murray20-cn}, and \ion{H}{1} column density $N_\mathrm{H\,I}$, in the high-latitude ($|b|>30\degree$) GALFA-\ion{H}{1} sky. Our main results are summarized as follows.
\begin{itemize}
    \item A strong positive $p_{353}$-$f_\mathrm{CNM}$ correlation is found in diffuse regions where there is no $p_{353}$-$N_\mathrm{H\,I}$ correlation. The $p_{353}$-$f_\mathrm{CNM}$ correlation behavior is column density-dependent, and transitions from a positive correlation to an anticorrelation at higher column densities ($N_\mathrm{H\,I}\gtrsim10^{21}\ \mathrm{cm^{-2}}$). 
    
    \item In the column density range $20.40<\log_{10}(N_{\mathrm{H\,I}}\  [\mathrm{cm}^{-2}])<20.60$, which spans $\sim1850~\rm deg^2$ of the GALFA-\ion{H}{1} sky, we find a consistent positive $p_{353}$-$f_\mathrm{CNM}$ correlation with Spearman correlation coefficient $\sim0.5$. {The column density range studied is motivated by excluding regions with low $f_{\rm CNM}$ dynamic range and significant molecular gas. A positive $p_{353}$-$f_\mathrm{CNM}$ correlation is found over a much larger column density range $\log_{10}(N_{\mathrm{H\,I}}\  [\mathrm{cm}^{-2}])<20.80$, spanning $\sim6000~\rm deg^2$ of the GALFA-\ion{H}{1} footprint.}
    
    \item We define simple models of magnetic field structure between phases. Fitting the models to data, we find that the observed positive $p_{353}$-$f_\mathrm{CNM}$ correlation is consistent with a higher degree of magnetic alignment in the CNM than in the WNM. On the other hand, the correlation does not vary significantly with dispersion in plane-of-sky polarization angle $\mathcal{S}$, nor with the number of \ion{H}{1} components along the LOS characterized by $N_c$, nor with the degree of alignment between \ion{H}{1} structures and the 353 GHz POS magnetic field orientation. 
    
    \item We find that a simple assumption of a disordered WNM-associated magnetic field relative to a more uniform CNM field is not sufficient to both explain the steep slope of the $p_{353}$-$f_\mathrm{CNM}$ relation and the observed maximum $p_{353}$. To explain the discrepancy, we hypothesize that an additional fraction of the non-CNM-associated dust column is also magnetically ordered, with the fraction of total ordered column proportional to the CNM fraction. Fixing $p_{\rm max}$ and fitting a two-parameter model to data results in the best-fit values of WNM order parameter $\kappa_{\mathrm{WNM}}=0.22$, and ratio of magnetically ordered regions $R_{\mathrm{ordered}}=4.68$. In other words, the dust column associated with the WNM has a mean polarization fraction that is 22\% of the dust column associated with the CNM, and in addition to the CNM, an additional column that corresponds to ~18.4\% of the HI mass is maximally polarized.
    
    \item The large best-fit value of $R_{\mathrm{ordered}}=4.68$ suggests that a significant fraction of the non-CNM column is also magnetically ordered relative to a disordered WNM column. The ratio translates to an average of 18.4\% of the total \ion{H}{1} mass having the same relatively low degree of magnetic field disorder as the CNM column, and we speculate that this corresponds to the UNM gas. 

    \item Our results showing $p_{353}$-$f_\mathrm{CNM}$ correlation and the CNM column being more magnetically ordered also have potential implications for dust grain models. If the observed maximally polarized regions are generally associated with high CNM content, and the CNM column contributes much more significantly than the WNM column to the LOS-averaged polarization fraction, then it is reasonable to expect minimal geometrical depolarization over the small CNM scale, resulting in high observed $p_{\rm max}$ in these regions.
\end{itemize}

Putting it all together, the observational constraints presented in this study are most consistent with the physical picture of a magnetically ordered CNM column forming out of a WNM with more disordered magnetic fields. This is a {new}, direct constraint on the mean degree of magnetic field disorder between neutral phases of the ISM.

\begin{figure*}[t]
    \centering
    \begin{minipage}{\linewidth}
        \centering
        \includegraphics[width=0.95\textwidth]{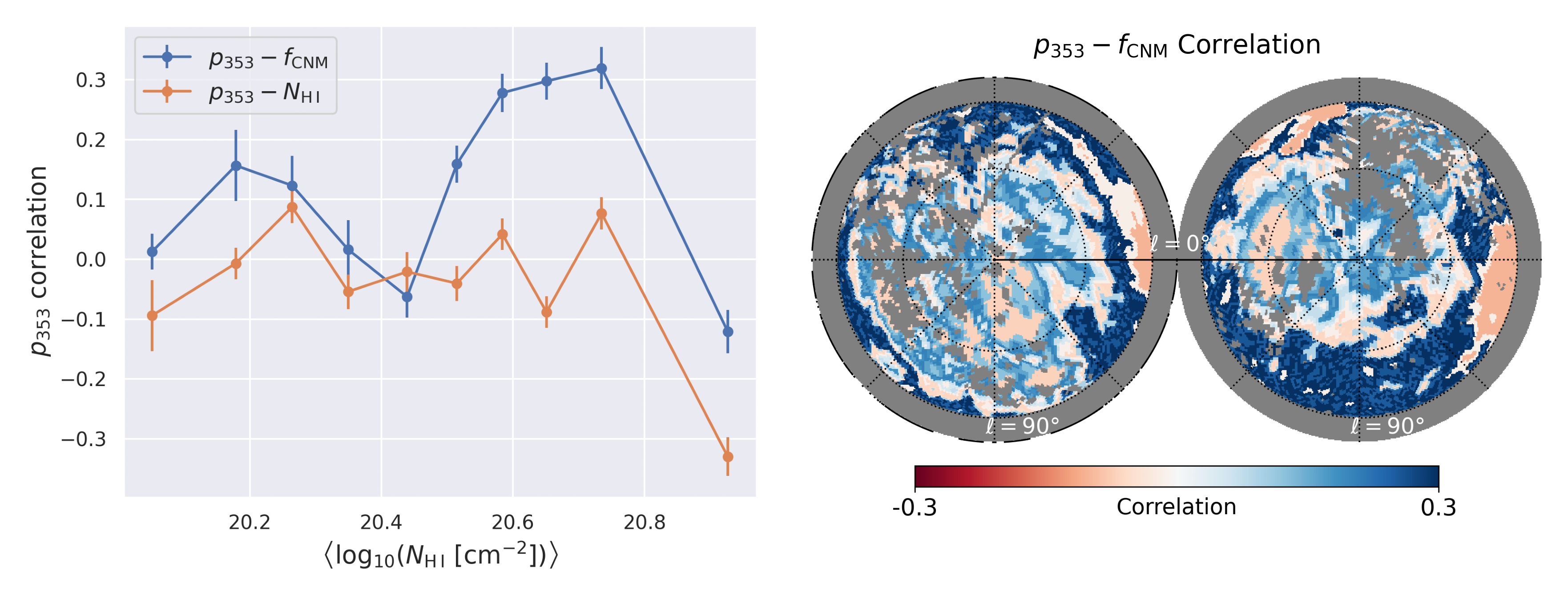}
        \caption{Column-density-binned $p_{353}$-$f_{\mathrm{CNM}}$ vs. $p_{353}$-$N_{\mathrm{H\,I}}$ correlation in the full high latitude ($|b|>30\degree$) sky using \ion{H}{1}4PI data. Left: Spearman correlation coefficients computed in equal sightline bins of $N_{\mathrm{H\,I}}$. Right: Map of correlation between $p_{353}$ and $f_{\mathrm{CNM}}$, created by dividing the \ion{H}{1}4PI sky into 20 equal sightline $N_{\mathrm{H\,I}}$ bins, colored by the $p_{353}$-$f_{\mathrm{CNM}}$ correlation coefficient value in each bin. The weaker positive $p_{353}$-$f_{\mathrm{CNM}}$ correlation compared to the GALFA-\ion{H}{1} version in Figure \ref{fig:p353_fcnm_corr} is likely attributable to the difference in $f_{\mathrm{CNM}}$ dynamic range between these two footprints as discussed in Appendix \ref{appx:galfa_v_hi4pi}. }
        \label{fig:p353_fcnm_corr_hi4pi}
    \end{minipage}
    \begin{minipage}{\linewidth}
        \centering
        \includegraphics[width=0.95\textwidth]{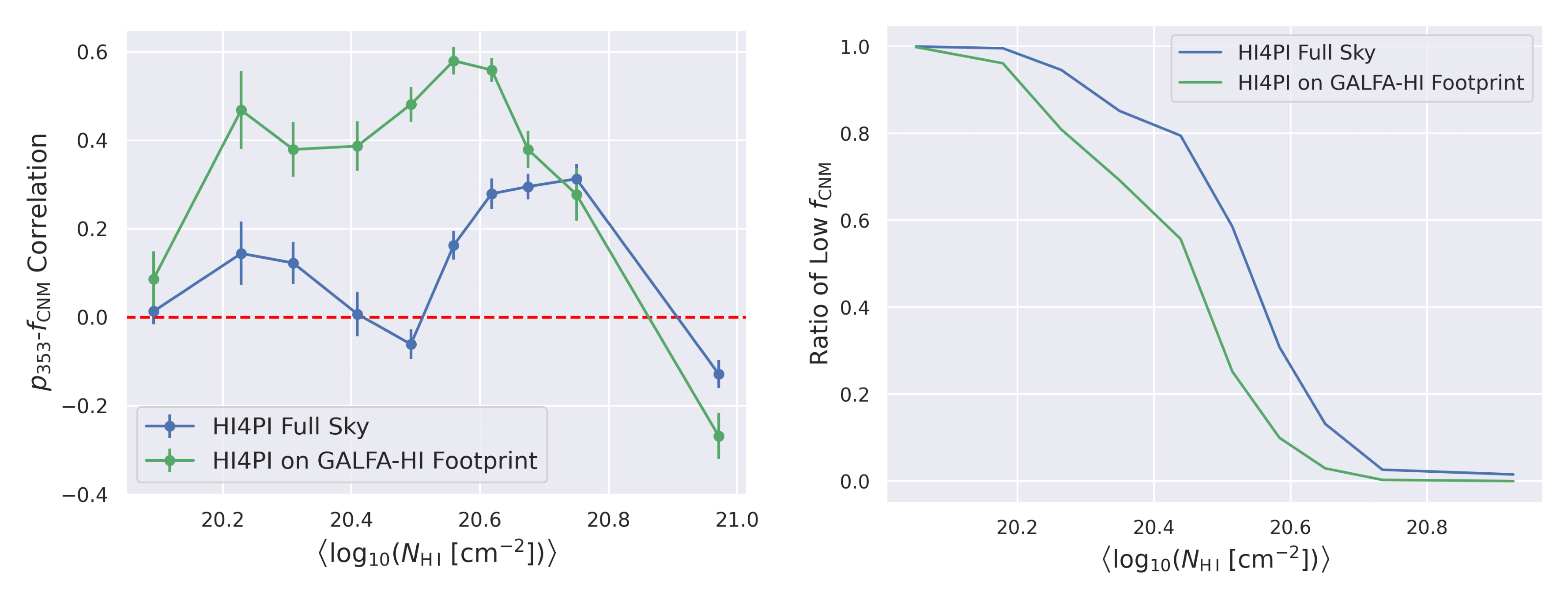}
        \caption{Left: Column-density-binned $p_{353}$-$f_{\mathrm{CNM}}$ vs. $p_{353}$-$N_{\mathrm{H\,I}}$ correlation in the high latitude ($|b|>30\degree$) sky using \ion{H}{1}4PI data within the GALFA-\ion{H}{1} footprint. The degree of correlation is comparable to the GALFA-\ion{H}{1} data version in Figure \ref{fig:p353_fcnm_corr}. Right: Ratio of sightlines with $f_{\mathrm{CNM}}<0.03$ across the same column density bins in the \ion{H}{1}4PI vs GALFA-\ion{H}{1} footprints. The GALFA-\ion{H}{1} footprint contains a significantly lower proportion of $f_{\mathrm{CNM}}\sim0$ sightlines except in the lowest and highest $N_{\mathrm{H\,I}}$ bins.}
        \label{fig:p353_fcnm_hi4pi2galfa}
    \end{minipage}
\end{figure*}

\section{Acknowledgments} We thank the anonymous referee for insightful comments that helped improve the paper. We thank Marc-Antoine Miville-Desch\^enes, Antoine Marchal, Peter Martin, Norm Murray, Gina Panopoulou, and Brandon Hensley for helpful discussions. This work was supported by the National Science Foundation under grant No. AST-2106607. 
This publication utilizes the Galactic ALFA HI (GALFA-\ion{H}{1}) survey data set obtained with the Arecibo $L$-band Feed Array (ALFA) on the Arecibo 305 m telescope. The Arecibo Observatory is operated by SRI International under a cooperative agreement with the National Science Foundation (AST-1100968), and in alliance with Ana G. M\'endez-Universidad Metropolitana and the Universities Space Research Association. The GALFA-\ion{H}{1} surveys have been funded by the NSF through grants to Columbia University, the University of Wisconsin, and the University of California. This paper also makes use of observations obtained with Planck, an ESA science mission, with instruments and contributions directly funded by ESA Member States, NASA, and Canada. The authors acknowledge Interstellar Institute’s program “II6” and the Paris-Saclay University’s Institut Pascal for hosting discussions that nourished the development of the ideas behind this work.

%

\vspace{5mm}


\software{Astropy \citep{astropy13, astropy18, astropy22},  
          NumPy \citep{numpy11},
          scipy \citep{scipy20},
          matplotlib \citep{matplotlib07},
          healpy \citep{healpy05-hp},
          PyMC \citep{salvatier16-mc}
          }



\appendix

\section{GALFA-HI and HI4PI Comparisons} \label{appx:galfa_v_hi4pi}
Here we extend the comparison of $p_{353}$-$N_{\mathrm{H\,I}}$ vs. $p_{353}$-$f_{\mathrm{CNM}}$ correlations in different column density regimes to the full high Galactic latitude ($|b|>30\degree$) sky using \ion{H}{1}4PI data. As described in Section \ref{subsec:hi_column}, \ion{H}{1}4PI covers the full sky at 16.2\arcmin\ angular resolution. We utilize \ion{H}{1}4PI column density and $f_{\mathrm{CNM}}$ maps smoothed to 80\arcmin\ to match the $p_{353}$ resolution \citep{hi4pi16-rl, hensley22-pa}. Figure \ref{fig:p353_fcnm_corr_hi4pi} shows the $p_{353}$-$N_{\mathrm{H\,I}}$ correlation coefficients in different column density regimes on the left, and the \ion{H}{1}4PI version of the correlation map produced by dividing the sky into 20 equal-number-of sightline regions on the right. The $p_{353}$-$N_{\mathrm{H\,I}}$ results in the $|b|>30\degree$ \ion{H}{1}4PI sky are consistent with our GALFA-\ion{H}{1} results and \citet{planck18-cr}, showing little correlation at lower column density, and an anticorrelation at high column densities. The correlation with $f_{\mathrm{CNM}}$ is also consistent with the GALFA-\ion{H}{1} trends, showing a positive correlation in column density regimes where there is no $p_{353}$-$N_{\mathrm{H\,I}}$ relation, and a transition to anticorrelation at higher column densities. However, compared to the GALFA-\ion{H}{1} result, the \ion{H}{1}4PI $p_{353}$-$N_{\mathrm{H\,I}}$ correlation is weaker and does not extend to lower column density bins $\log_{10}(N_{\mathrm{H\,I}}\  [\mathrm{cm}^{-2}])<20.4$. While the qualitative behavior of the $p_{353}$-$f_{\rm CNM}$ relation between datasets is consistent, we explore the possible explanations for the difference in the degree of correlation. 

First, since both GALFA-\ion{H}{1} and \ion{H}{1}4PI are smoothed to 80\arcmin\ to match the $p_{353}$ resolution, the difference in their native angular resolution should not affect the $p_{353}$-$f_{\rm CNM}$ correlation results. Furthermore, when comparing the \ion{H}{1}4PI and GALFA-\ion{H}{1} versions of $f_\mathrm{CNM}$ maps, \citet{hensley22-pa} found excellent agreement in overlapping regions, with a Spearman correlation coefficient of 0.97 when both maps are smoothed to the same resolution. To determine whether the differences are attributable to the data sets or to the sky areas considered, we analyze HI4PI data on the GALFA-HI footprint. Figure \ref{fig:p353_fcnm_hi4pi2galfa} shows the comparison between the correlation using the full \ion{H}{1}4PI data vs. the region that overlaps GALFA-\ion{H}{1}. The $p_{353}$-$f_{\mathrm{CNM}}$ correlation at lower column density regimes $\log_{10}(N_{\mathrm{H\,I}}\  [\mathrm{cm}^{-2}])<20.6$ is stronger for the GALFA-\ion{H}{1}-overlapped region than the full \ion{H}{1}4PI data sky, and is consistent with the GALFA-\ion{H}{1} results in Figure \ref{fig:p353_fcnm_corr}. This implies that a variation in the $f_{\mathrm{CNM}}$ distribution between GALFA-\ion{H}{1} and \ion{H}{1}4PI footprints might play a larger role than any differences between these two datasets.

We already considered the effect of $f_\mathrm{CNM}$ distribution when comparing $p_{353}$-$f_{\mathrm{CNM}}$ distribution in different column density bins in Figure \ref{fig:compare_col_den}. In particular, 80\% of low column density regions with $\log_{10}(N_{\mathrm{H\,I}}\  [\mathrm{cm}^{-2}])<20.40$ has $f_\mathrm{CNM}<0.03$. {When the $f_{\rm CNM}$ dynamic range is limited, the general scatter in $p_{353}$ is significant compared to the $p_{353}$ variation driven by $f_\mathrm{CNM}$, potentially diluting any $p_{353}$-$f_{\rm CNM}$ correlation. From the \ion{H}{1}4PI version of the $f_\mathrm{CNM}$ map, $\sim57\%$ of the sightlines at high latitude have $f_\mathrm{CNM} < 0.03$, while only $\sim44\%$ of the sightlines in the GALFA-\ion{H}{1} footprint satisfy this condition. When we adopt a column density cut of $20.6<\log_{10}(N_{\mathrm{H\,I}}\  [\mathrm{cm}^{-2}])<20.8$ on the \ion{H}{1}4PI dataset so that it has similar $f_{\rm CNM}$ dynamic range as the GALFA-\ion{H}{1} region of interest analyzed in the paper, and fit model2 to the new region, we find that the best-fit model parameters are qualitatively consistent with the GALFA-\ion{H}{1} results presented in Section \ref{subsec:phase_coh_imp}. The best-fit $\kappa_{\rm WNM}=0.23^{+0.01}_{-0.02}$ agrees with the GALFA-\ion{H}{1} result to within uncertainty, while the best-fit $R_{\rm ordered}=3.2^{+0.20}_{-0.22}$ is lower than the GALFA-\ion{H}{1} result of $4.68^{+0.21}_{-0.20}$ but still significantly larger than 1, pointing to the same picture that a significant fraction of the non-CNM is also magnetically-ordered. Thus, while there is likely some spatial variation of the parameters driving the $p_{353}$-$f_\mathrm{CNM}$ relation, the general picture of a more ordered CNM column relative to a disordered WNM column still applies, and a major factor in the difference between the GALFA-\ion{H}{1} and \ion{H}{1}4PI datasets is the more limited $f_\mathrm{CNM}$ dynamic range and SNR in the full \ion{H}{1}4PI footprint. Within the GALFA-\ion{H}{1} footprint presented in the main analysis, there is no strong evidence of spatial variation. When we repeat our analysis on two halves of the dataset randomly drawn from the full set, the best-fit $\kappa$ and $R_{\rm ordered}$ parameters agree between those samples within uncertainties. Future studies can more quantitatively examine the spatial variation of the $p_{353}$-$f_\mathrm{CNM}$ correlation by employing an improved CNM fraction map with higher SNR at low $f_\mathrm{CNM}$ sightlines, and considering a more complete dust polarization model that explicitly models other contributions to $p_{353}$ scatter such as grain properties and inclination angle in regions where $f_\mathrm{CNM}\sim~0$.}


\onecolumngrid

\bibliography{bcoherence}{}
\bibliographystyle{aasjournal}



\end{document}